\documentclass[%
 aip,
 amsmath,amssymb,
reprint,%
]{revtex4-1}

\usepackage{graphicx}
\usepackage{dcolumn}
\usepackage{bm}

\usepackage[utf8]{inputenc}
\usepackage[T1]{fontenc}
\usepackage{mathptmx}

\usepackage{color}

\newcommand{\stkout}[1]{\ifmmode\text{\sout{\ensuremath{#1}}}\else\sout{#1}\fi}

\graphicspath{{img/}}

\begin{document}

\title{
Predominance of non-adiabatic effects in zero-point renormalization of the electronic band gap.
}

\author{Anna Miglio}
\thanks{Anna Miglio and V\'eronique Brousseau-Couture contributed equally to this work.}
\affiliation{Institute of Condensed Matter and Nanosciences, UCLouvain,
B-1348 Louvain-la-Neuve, Belgium}

\author{V\'eronique Brousseau-Couture}
\thanks{Anna Miglio and V\'eronique Brousseau-Couture contributed equally to this work.}
\affiliation{
D\'epartement de Physique et Regroupement Qu\'eb\'ecois sur les Mat\'eriaux de Pointe, Universit\'e de Montreal, C.P. 6128, Succursale Centre-Ville, Montreal, Canada H3C 3J7
}

\author{Emile Godbout}
\affiliation{
D\'epartement de Physique et Regroupement Qu\'eb\'ecois sur les Mat\'eriaux de Pointe, Universit\'e de Montreal, C.P. 6128, Succursale Centre-Ville, Montreal, Canada H3C 3J7
}

\author{Gabriel Antonius }
\affiliation{Department of Physics, University of California at Berkeley, California 94720, USA
and Materials Sciences Division, Lawrence Berkeley National Laboratory, Berkeley, California 94720, USA}
\affiliation{
D\'epartement de Chimie, Biochimie et Physique, Institut de recherche sur l'hydrog\`ene, Universit\'e du Qu\'ebec \`a Trois-Rivi\`eres, Trois-Rivi\`eres, Canada}

\author{Yang-Hao Chan}
\affiliation{Department of Physics, University of California at Berkeley, California 94720, USA
and Materials Sciences Division, Lawrence Berkeley National Laboratory, Berkeley, California 94720, USA}

\author{Steven G. Louie}
\affiliation{Department of Physics, University of California at Berkeley, California 94720, USA
and Materials Sciences Division, Lawrence Berkeley National Laboratory, Berkeley, California 94720, USA}

\author{Michel C\^ot\'e}
\affiliation{
D\'epartement de Physique et Regroupement Qu\'eb\'ecois sur les Mat\'eriaux de Pointe, Universit\'e de Montreal, C.P. 6128, Succursale Centre-Ville, Montreal, Canada H3C 3J7
}

\author{Matteo Giantomassi }
\affiliation{Institute of Condensed Matter and Nanosciences, UCLouvain, B-1348 Louvain-la-Neuve, Belgium}

\author{ Xavier Gonze$^*$ }
\email{xavier.gonze@uclouvain.be}
\affiliation{Institute of Condensed Matter and Nanosciences, UCLouvain, B-1348 Louvain-la-Neuve, Belgium}
\affiliation{Skolkovo Institute of
Science and Technology, Skolkovo Innovation Center, Nobel St. 3, Moscow, 143026, Russia.}

\date{\today}

\begin{abstract}
Electronic and optical properties of materials are affected by atomic motion through the electron-phonon interaction:
not only band gaps change with temperature,
but even at absolute zero temperature, zero-point motion causes band-gap renormalization.
We present a large-scale
first-principles evaluation of the zero-point renormalization of band edges beyond the adiabatic approximation.
For materials with light elements, the band gap renormalization is often larger than 0.3 eV, and up to 0.7 eV.
This effect cannot be ignored if accurate band gaps are sought.
For infrared-active materials, global agreement with available experimental data is obtained only when non-adiabatic effects
are taken into account. They even dominate zero-point renormalization for many materials,
as shown by a generalized Fr\"ohlich model
that includes multiple phonon branches, anisotropic and degenerate electronic extrema,
whose range of validity is established by comparison with first-principles results.\\\\
\emph{This is a post-peer-review, pre-copy-edited
version of the article published in npj Computational Materials (2020) 6:167. The final authenticated version is available online
at: http://dx.doi.org/10.1038/s41524-020-00434-z \\
Note that the supplementary material, included in the main file, is structured differently than in the published version, with a different numbering scheme.}
\end{abstract}

\maketitle


\label{sec:Intro}

The electronic band gap is arguably the most important characteristic of semiconductors and insulators.
It determines optical and luminescent thresholds,
but is also a prerequisite for characterizing band offsets at interfaces and deep electronic levels created by defects~\cite{Yu2010}.
However, accurate band gap computation is a challenging task.
Indeed, the vast majority of first-principles calculations relies on Kohn-Sham Density-Functional Theory (KS-DFT), valid for ground state properties~\cite{Martin2004}, that
delivers a theoretically unjustified value of the band gap in the standard approach, even with exact KS potential~\cite{Perdew1983,Sham1983,Martin2016}.

The breakthrough came from many-body perturbation theory, with the so-called
GW approximation, first non-self-consistent (G$_0$W$_0$) by Hybertsen and Louie in 1986~\cite{Hybertsen1986},
then twenty years later self-consistent (GW) \cite{Schilfgaarde2006} and further improved by accurate vertex corrections from electron-hole excitations (GWeh) \cite{Shishkin2007}. The latter
methodology, at the forefront for band-gap computations,
delivers a 2\%-10\%
accuracy, usually overestimating the experimental band gap. GWeh calculations are computationally very demanding, typically about two orders of magnitude more than G$_0$W$_0$, itself two orders of magnitude more time-consuming than KS-DFT calculations, roughly speaking.

Despite being state-of-the-art, {\it such studies ignored completely the electron-phonon interaction}.
The {\color{black} electron-phonon interaction} drives most of the temperature dependence of the electronic structure of semiconductors and insulators,
but also yields a zero-point motion gap modification at $T=0$~K,
often termed zero-point renormalization of the gap (ZPR$_{\rm g}$)  for historical reasons.

\begin{figure}
\includegraphics[width=0.40\textwidth]{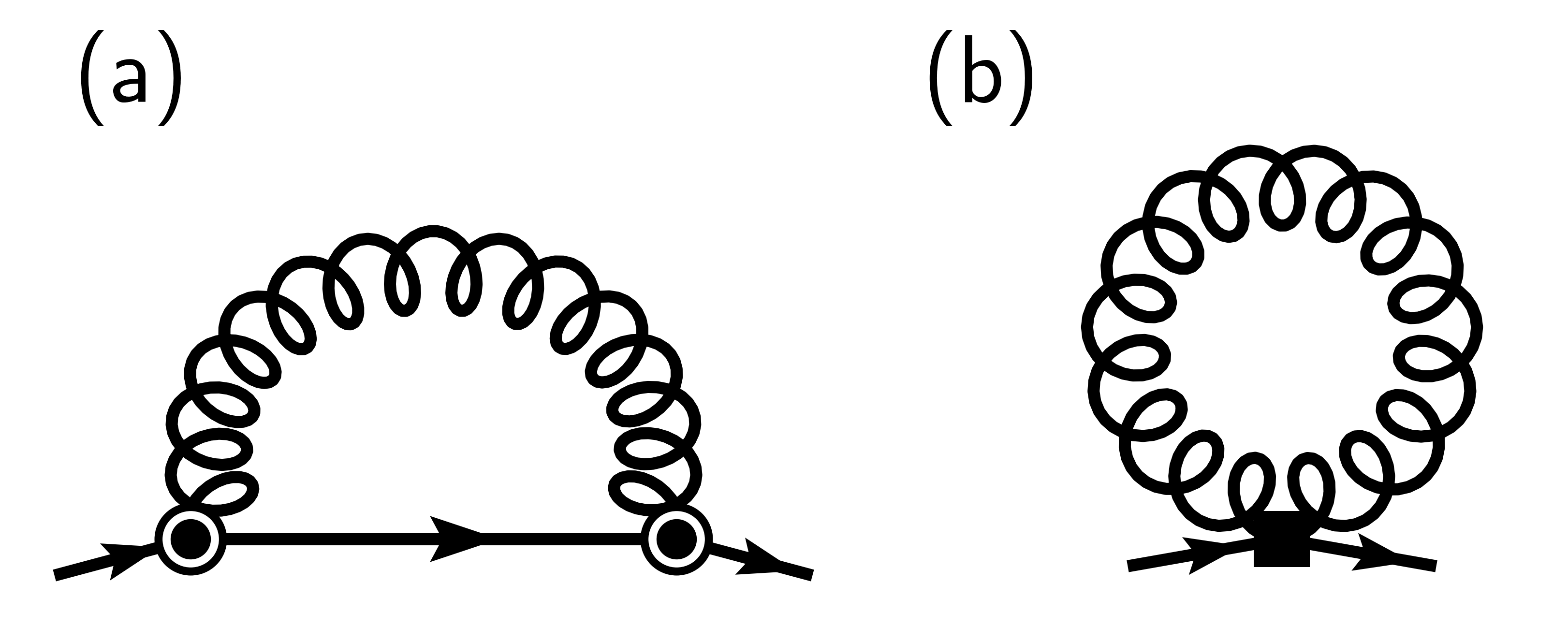}
\caption{ \label{fig:FanDW}
Lowest-order diagrams for the contribution of the electron-phonon interaction to the zero-point renormalization of the gap ZPR$_{\rm g}$. (a) Fan diagram, with two first-order screened {\color{black} electron-phonon} vertices; (b) Debye-Waller diagram, with one second-order screened {\color{black} electron-phonon interaction} vertex.
}
\end{figure}

The ZPR$_{\rm g}$ had been examined 40 years ago, by Allen, Heine and Cardona (AHC)~\cite{Allen1976, Allen1981}
who clarified the early theories by Fan~\cite{Fan1951} and Antoncik~\cite{Antoncik1955}.
Their approach is, like the GW approximation, rooted in many-body perturbation theory, where, at the lowest order,
two diagrams contribute, see Fig.~\ref{fig:FanDW}, the so-called ``Fan" diagram, with two 1$^{st}$-order electron-phonon vertices and the ``{\color{black} Debye-Waller}" diagram, with one 2$^{nd}$-order electron-phonon vertex.
In the context of semi-empirical calculations, the AHC method was applied to Si and Ge,
introducing the adiabatic approximation, in which the phonon frequencies are neglected with respect to electronic eigenenergy differences and
replaced by a small but non-vanishing imaginary broadening~\cite{Allen1981}.
It was later extended without caution
to GaAs and a few other III-V semiconductors~\cite{Kim1986,Zollner1991}.

In this work, we present first-principles AHC ZPR$_{\rm g}$ calculations {\it beyond the adiabatic approximation, for 30 materials}.
Comparing with experimental band gaps,
we show that adding ZPR$_{\rm g}$ improves the GWeh first-principles band gap, and moreover, that
 {\it the ZPR$_{\rm g}$ has the same order of magnitude as the G$_0$W$_0$
to GWeh correction} for half of the materials (typically materials with light atoms, e.g. O, N ...) on which GWeh has been tested.
Hence, the GWeh level of sophistication misses its target for many materials with light atoms, if the ZPR$_{\rm g}$ is not taken into account. With it, {\it the theoretical agreement with direct measurements of experimental ZPR$_{\rm g}$ is improved}. This also demonstrate the {\it crucial importance of phonon dynamics} to reach this level of accuracy.

Indeed, first-principles calculations of the ZPR$_{\rm g}$ using the AHC theory are very challenging,
and only started one decade ago, on a case-by-case basis (see Refs~\cite{Marini2008,Giustino2010,
Cannuccia2011,Gonze2011a,Cannuccia2012,Kawai2014,Ponce2014a,Ponce2014b,
Antonius2015,Ponce2015,Nery2018,
note-additionalrefs}, and the SI II.D), usually relying on the adiabatic approximation, and without comparison with experimental data.
An approach to the ZPR$_{\rm g}$, alternative to the AHC one, relies on computations of the band gap at fixed, distorted geometries, for large supercells (see Refs.)~\cite{Capaz2005,Gonze2011a,
Patrick2013,Monserrat2013,Antonius2014,Monserrat2014a,Monserrat2014b,Monserrat2015,Engel2016} and the SI~\cite{note-additionalrefs}, Sec.II.D).
As the adiabatic approximation is inherent in this approach, we denote it as ASC, for ``adiabatic supercell".
A recent publication by Karsai and co-workers~\cite{Karsai2018} presents ASC ZPR$_{\rm g}$
based on DFT values as well as based on G$_0$W$_0$ values for 18 semiconductors, with experimental comparison for 9 of them.
Both AHC and ASC methodologies have been recently reviewed~\cite{Giustino2017}.

Although the adiabatic approximation had already been criticized by Ponc\'e {\it et al.} in 2015~\cite{Ponce2015} (SI Sec.III.A) as causing an unphysical divergence of the adiabatic AHC expression for infrared-active materials with vanishing imaginary broadening, thus invalidating all {\it adiabatic AHC}
calculations except for non-{\color{black} infrared-}active materials, the full consequences of the adiabatic approximation have not yet been recognized {\it for ASC}.
We will show that the non-adiabatic AHC approach
outperforms the ASC approach, so that the predictions arising from the mechanism by which the ASC approach bypass the adiabatic AHC divergence
problem for infrared-active materials are questionable.
This is made clear by a generalized Fr\"ohlich model with a
few physical parameters, that can be determined either from first principles or from experimental data.

Although the full electronic and phonon band structures do not enter in this model,
and the {\color{black} Debye-Waller} diagram is ignored, for many materials it accounts for more than half the ZPR$_{\rm g}$ of the full first-principles non-adiabatic AHC ZPR$_{\rm g}$.
As this model depends crucially on non-adiabatic effects, it demonstrates the failure of the adiabatic hypothesis be it for the AHC or the ASC approach.

By the same token, we
also show the domain of validity and accuracy of model Fr\"ohlich large-polaron calculations based on the continuum hypothesis, that have been the subject of decades of research~\cite{Froehlich1954,Feynman1955,
Mishchenko2000,Mishchenko2003,
Devreese2009,Emin2012,
Hahn2018a}. Such model Fr\"ohlich Hamiltonian
capture well the ZPR for about half of the materials in our list, characterized by their strong {\color{black} infrared} activity,
while it becomes less and less adequate for
decreasing ionicity.
In the present context, the Fr\"ohlich large-polaron model provides an intuitive picture of the physics of the ZPR$_{\rm g}$.

\section*{Results}
\label{sec:Results}
\noindent
{\bf{Zero-point renormalization: experiment versus first principles.}}
\label{sec:ZPR_Exp_FP}

\begin{figure}
\includegraphics[width=0.48\textwidth]{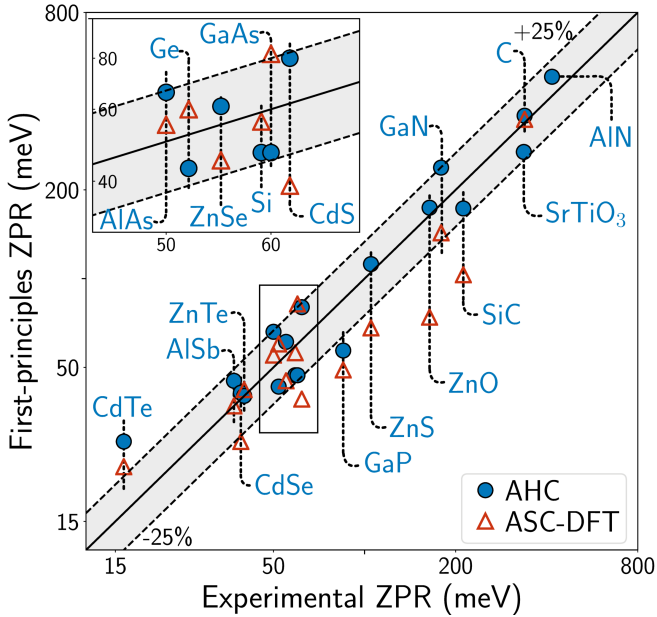}
\caption{ \label{fig:FPvsExp}
Absolute values of first-principles band gap renormalization ZPR$_{\rm g}$
compared with experimental ones.
Blue full circles: present calculations,
using non-adiabatic AHC, based on KS-DFT ingredients;
red empty triangles: adiabatic supercell KS-DFT  
Dashed lines: limits at which the smallest of both ZPR is 25\% smaller than the largest one (note that the scales are logarithmic).
For consistency, the computed lattice ZPR$^{\rm lat}_{\rm g}$ were added to both ASC and AHC.
The majority of non-adiabatic AHC-based results fall 
See numerical values in Table~S2
of the SI.}
\end{figure}

Fig.~\ref{fig:FPvsExp} compares first-principles ZPR$_{\rm g}$ with experimental values.
As described in the METHODS section, and in SI Sec.II.C, the correction due to zero-point motion effect on the lattice parameter{\color{black}, ZPR$^{\rm lat}_{\rm g}$,} has been added to fixed volume results from both
non-adiabatic AHC (present calculations) and ASC methodologies~\cite{Karsai2018}.
While for a few materials experimental ZPR$_{\rm g}$ values
are well established, within 5-10\%,
globally, experimental uncertainty is larger, and can
hardly be claimed to be better than 25\%
for the majority of materials, see Sec.II.A
in the SI.
This will be our tolerance.

Let us focus first on the ASC-based results.
For the 16 materials present in both Ref.~\onlinecite{Karsai2018} and the experimental set described in the SI Table~S1, the ASC vs experimental discrepancy is more than 25\%
for more than half of the materials
~\cite{note-Karsai}.
There is a global trend to underestimation by ASC, although CdTe is overestimated.

By contrast, the non-adiabatic AHC ZPR$_{\rm g}$ (blue full circles) and experimental ZPR$_{\rm g}$ agree with each other within 25\% for 16 out of the 18 materials.
The outliers are CdTe with a 43\% overestimation
by theory and GaP with a 33\% underestimation. For none of these the discrepancy is a factor of two or larger.
On the contrary, from the ASC approaches, several materials show underestimation of the ZPR$_{\rm g}$ by more than a factor of 2. The materials showing such large underestimation (CdS, ZnO, SiC) are all quite ionic, while more covalent materials (C, Si, Ge, AlSb, AlAs) are better described.

Therefore, Fig.~\ref{fig:FPvsExp} clearly shows that {\it the non-adiabatic AHC approach performs significantly better than the ASC approach}.
AHC ZPR$_{\rm g}$ and ASC ZPR$_{\rm g}$ also differ by more than a factor of two for TiO$_2$ and MgO (see SI Sec.II.D), although no experimental ZPR$_{\rm g}$ is available
for these materials to our knowledge.

\begin{figure}
\includegraphics[width=0.45\textwidth]{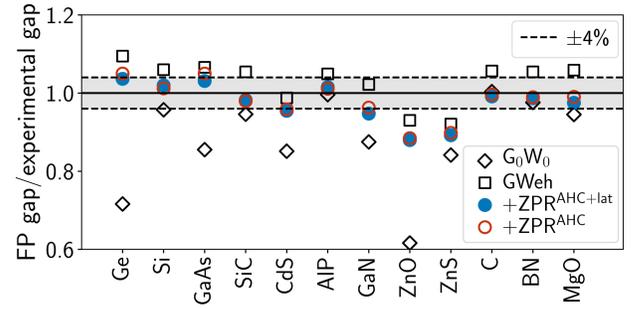}
\caption{\label{fig:GapExpFP-lat}
Ratio between first-principles band gaps and experimental ones~\cite{Cardona2005}.
First-principles results without electron-phonon interation\cite{Shishkin2007} are based on G$_0$W$_0$ (diamonds) or on GWeh (squares).
Blue full circles are the GWeh results to which AHC ZPR$_{\rm g}$ and lattice ZPR$^{\rm lat}_{\rm g}$ have been added, while empty red circles are the GWeh results to with only AHC ZPR$_{\rm g}$ was added.
GWeh usually overshoots, while adding the electron-phonon interaction gives better global agreement.
See numerical values in Table~S3
of the SI.
}
\end{figure}

We now examine band gaps.  Fig.~\ref{fig:GapExpFP-lat} presents the ratio between first-principles band gaps and corresponding experimental values,
for 12 materials. The best first-principles values at fixed equilibrium atomic position, from GWeh \cite{Shishkin2007}, are represented, as well as their non-adiabatic AHC ZPR corrected values.

For GWeh without ZPR$_{\rm g}$, a 4\% agreement is
obtained only for two materials (CdS and GaN).
There is indeed a clear, albeit small, tendency of GWeh to overestimate
the band gap value, except for the 3 materials containing shallow core d-electrons (ZnO, ZnS and CdS) that are underestimated.
By contrast, if the non-adiabatic AHC ZPR$_{\rm g}$ is added to the GWeh data (blue dots), a 4\% agreement is obtained for 9 out of the 12 materials (8 if ZPR$^{\rm lat}_{\rm g}$ is not included).

For ZnO and ZnS, with a final 10-12\% underestimation,
and CdS with a 5\% underestimation,
we question the GWeh ability to produce accurate fixed-geometry band gaps at the level obtained for the other
materials, due to the presence of rather localized 3d electrons in Zn
and 4d electrons in CdS.

As a final lesson from Fig.~\ref{fig:GapExpFP-lat}, we note that
for 4 out of the 12 materials (SiC, AlP, C, and
BN),
the ZPR$_{\rm g}$ is of similar size than the G$_0$W$_0$ to GWeh correction, and it is a significant fraction of it also for Si, GaN and MgO.
As mentioned earlier, GWeh calculations are much more time-consuming than G$_0$W$_0$ calculations, possibly even more time-consuming
than ZPR calculations (although we have not attempted to make a fair comparison).
Thus, for materials containing light elements, first row and likely second row (e.g. AlP) in the periodic table,
{\it GWeh calculations miss their target if not accompanied
by ZPR$_{\rm g}$ calculations}.
A review of the variance and accuracy of G$_0$W$_0$ calculations for these materials is discussed in Ref.~\onlinecite{Rangel2020}.

Table~S5
of the SI gathers our full set of ZPR results, beyond those present in the ASC or experimental sets.
It also includes
10 oxides,
while the experimental set only includes three materials containing oxygen (ZnO, MgO and SrTiO$_3$),
and there are none in Karsai's ASC set.
The ZPR$_{\rm g}$ of the band gap for materials containing light elements (O or lighter) is between \mbox{-157 meV} (ZnO) and \mbox{-699 meV} (BeO), while, relatively to the experimental band gap, it ranges from \mbox{-4.6\%} (ZnO) to \mbox{-10.8\%} (TiO$_2$-t).
This can hardly be ignored in accurate calculations of the gap.

\vspace{12 pt}

\noindent{\bf{Generalized Fr\"ohlich model.}}
\label{sec:gFr}

We now come back to the physics from which the ZPR$_{\rm g}$ originates, and explain our earlier observation that
the ASC describes reasonably well the more covalent materials, but can fail badly for ionic materials.
We argue that, for many polar materials, the ZPR$_{\rm g}$ is dominated by the diverging {\color{black} electron-phonon interaction} between zone-center LO phonons and electrons close to the band edges,
and {\it the slow (non-adiabatic) response of the latter}:
the effects due to comparatively fast phonons are crucial.
This was already the message from Fr\"ohlich~\cite{Froehlich1954} and Feynman~\cite{Feynman1955}, decades ago, initiating large-polaron studies.
However, large-scale
assessment of the adequacy of Fr\"ohlich model for real materials is lacking.
Indeed, the available flavors of
Fr\"ohlich model, based on continuum (macroscopic) electrostatic interactions,
do not cover altogether degenerate and anisotropic electronic band extrema
or are restricted to only one phonon branch, unlike most real materials~\cite{Mahan1965, Trebin1975,Devreese2010,Schlipf2018}.

In this respect, we introduce now a generalized Fr\"ohlich
model, gFr, whose form is deduced from first-principles equations in the long-wavelength limit (continuum approximation). Such model covers all situations and still uses as input only long-wavelength
parameters, that
can be determined either from first principles (SI Sec.IV.D)
or from experiment (SI Sec.IV.E).
We then find the corresponding ZPR$_{\rm g}$
from perturbation theory at the lowest order, evaluate it for our set of materials, and compares
its results to the ZPR$_{\rm g}$ obtained from full first-principles
computations.
The corresponding Hamiltonian writes (see the SI Sec.IV.B for detailed explanations):
\begin{equation}
\hat{H}^{\rm gFr}=\hat{H}^{\rm gFr}_{\rm el}+\hat{H}^{\rm gFr}_{\rm ph}+\hat{H}^{\rm gFr}_{\rm EPI} \, ,
\label{eq:HgFr}
\end{equation}
with {\color{black} \it (i)} an electronic part
{\color{black}
\begin{equation}
\hat{H}^{\rm gFr}_{\rm el}= \sum_{\mathbf{k}n}
\frac{k^2}{2m^*_n(\hat{\mathbf{k}})}
\hat{c}^+_{\mathbf{k}n}\hat{c}_{\mathbf{k}n} \, ,
\label{eq:HgFrel}
\end{equation}
}that includes
direction-dependent effective masses {\color{black} $m^*_n(\hat{\mathbf{k}})$},
governed by so-called Luttinger parameters in case of degeneracy,
{\color{black}electronic creation and annihilation operators,
$\hat{c}^+_{\mathbf{k}n}$
and
$\hat{c}_{\mathbf{k}n}$,
with $\mathbf{k}$ the electron wavevector
and $n$ the band index,
}
{\color{black} \it (ii)} the multi-branch phonon part,
{\color{black}
\begin{equation}
\hat{H}^{\rm gFr}_{\rm ph}= \sum_{\mathbf{q}j} \omega_{j0}(\hat{\mathbf{q}})
\hat{a}^+_{\mathbf{q}j}\hat{a}_{\mathbf{q}j} \, ,
\label{eq:HgFrph}
\end{equation}
with
direction-dependent phonon frequencies
$\omega_{j0}(\hat{\mathbf{q}})$,
phonon creation and annihilation operators,
$\hat{a}^+_{\mathbf{q}j}$
and
$\hat{a}_{\mathbf{q}j}$,
with $\mathbf{q}$ the phonon wavevector and $j$ the branch index,
}and {\color{black}finally \it (iii)} the electron-phonon interaction part
\begin{equation}
\hat{H}^{\rm gFr}_{\rm EPI}= \sum_{\mathbf{q}j,\mathbf{k}n'n}
g^{\rm gFr}(\mathbf{q}j,\mathbf{k}n'n)
\hat{c}^+_{\mathbf{k}+\mathbf{q}n'}\hat{c}_{\mathbf{k}n}
(\hat{a}_{\mathbf{q}j}+\hat{a}^+_{-\mathbf{q}j}) \, .
\label{eq:HgFrEPI}
\end{equation}
The $\mathbf{k}$ and $\mathbf{q}$ sums run over the
Brillouin zone.
The sum over $n$ and $n'$
runs only over the bands that connect to the degenerate
extremum{\color{black}, renumbered from 1 to $n_{\rm deg}$}.
The generalized Fr\"ohlich {\color{black} electron-phonon interaction}~\cite{Vogl1976,Verdi2015} is
\begin{eqnarray}
g^{\rm gFr}(\mathbf{q}j,\mathbf{k}n'n)=
\frac{i}{q} &&\frac{4\pi}{\Omega_0}
\Big(  \frac{1}{2\omega_{j0}(\hat{\mathbf{q}})V_{\rm BvK} }
\Big)^{1/2}
\frac{\hat{\mathbf{q}}.\mathbf{p}_j(\hat{\mathbf{q}})}
        {\epsilon^\infty(\hat{\mathbf{q}})}
\nonumber
\\
\times &&
\sum_m
s_{n'm}(\hat{\mathbf{k}}')
(s_{nm}(\hat{\mathbf{k}}))^*.
\label{eq:SM_ggFr}
\end{eqnarray}
This expression depends on the directions
$\hat{\mathbf{k}}$,
$\hat{\mathbf{q}}$,
and $\hat{\mathbf{k}}'$
($\mathbf{k}'=\mathbf{k}+\mathbf{q}$), but
{\it not explicitly on their norm} (hence only long-wavelength
parameters are used), except for the $\frac{1}{q}$ factor.
The {\color{black} electron-phonon} part also depends only on few
quantities: the
Born effective charges (entering the mode-polarity vectors $\mathbf{p}_j$), the macroscopic dielectric tensor $\epsilon^\infty$, and the phonon frequencies $\omega_{j0}${\color{black}, the primitive cell volume $\Omega_0$},
the Born-von Karman normalisation volume $V_{\rm BvK}$ corresponding to the $\mathbf{k}$ and $\mathbf{q}$ samplings.
The $s$ tensors
are symmetry-dependent unitary matrices, similar to
spherical harmonics.
Eqs.~(\ref{eq:HgFr}-\ref{eq:SM_ggFr}) define our generalized Fr\"ohlich Hamiltonian.
Although we will focus on its $T=0$ properties within perturbation theory, such Hamiltonian could be studied for many different purposes (non-zero T, mobility, optical responses ...),
like the original Fr\" ohlich model, for representative materials using first-principles or experimental parameters.

The corresponding ZPR$^{\rm gFr}_{\rm c}$ can be obtained with a perturbation treatment (SI Sec.IV.C), giving
\begin{eqnarray}
{\rm ZPR}_{\rm c}^{\rm gFr}&=&
-\sum_{jn}
\frac{1}{\sqrt{2}\Omega_0
{\color{black}n_{\rm deg}}}
\int_{4\pi} d\hat{\mathbf{q}}
\big( m_n^*(\hat{\mathbf{q}})
\big)^{1/2}
\nonumber
\\
&\times &\big(\omega_{j0}(\hat{\mathbf{q}})
\big)^{-3/2}
\Big(
\frac{\hat{\mathbf{q}}.\mathbf{p}_j(\hat{\mathbf{q}})}
        {\epsilon^\infty(\hat{\mathbf{q}})}
\Big)^2.
\label{eq:ZPR_c_Fr}
\end{eqnarray}

A similar expression exists for the valence ZPR$^{\rm gFr}_{\rm v}$.
The few material parameters needed in Eq.~(\ref{eq:ZPR_c_Fr}) can be obtained from
experimental measurements, but are most easily computed from first principles,
using density-functional perturbation theory
{\it with calculations only at $\mathbf{q}=\Gamma$} (e.g. no phonon band structure calculation).
Eq.~(\ref{eq:ZPR_c_Fr}) can be evaluated for all band extrema in our set of materials, irrespective of whether the extrema are located at $\Gamma$ or other
points in the Brillouin Zone (e.g. X for the valence band of many oxides, with anisotropic effective mass),
whether they are degenerate (e.g. the three-fold degeneracy of the top of the valence band of many III-V or II-VI compounds),
and irrespective of the number of phonon branches
(e.g. 3 different LO frequencies for TiO$_2$, moreover varying with the direction along which \mbox{$\mathbf{q}$ $\rightarrow 0$}).

\begin{figure}
\includegraphics[width=0.45\textwidth]{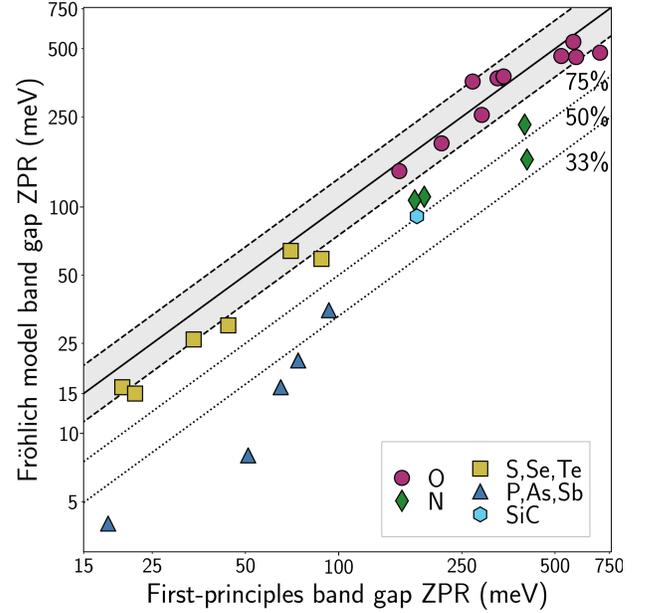}
\caption{\label{fig:FROvsFPb_gap_fourth}
Zero-point renormalization of the band gap: comparison between first-principles AHC values and the values obtained from the generalized Fr\"ohlich model fed with first-principles values, see Eq.~(\ref{eq:ZPR_c_Fr}),  for {27} materials (non-ionic group IV materials are omitted).
The markers identify materials of similar ionicity: oxides (purple circles), other II-VI materials containing S, Se or Te (yellow squares), nitrides (green diamonds), other III-V materials, containing P, As, or Sb (dark blue triangles), and group IV material SiC (light blue hexagon).
Dashed lines: limits at which the smallest of both ZPR is 25\% smaller than the largest one.
Dotted lines: limits at which the Fröhlich model ZPR misses respectively 50\% and 66\% of the AHC values (note that the scales are logarithmic).
See numerical values in Table~S5
of the SI.}
\end{figure}

Fig.~\ref{fig:FROvsFPb_gap_fourth} compares the band gap ZPR from the first-principles non-adiabatic AHC methodology and from the generalized Fr\"ohlich model.

The 30 materials can be grouped into five sets, based on their ionicity:
11 materials containing oxygen, rather ionic, for which the Born effective charges and the ZPR are quite large,
6 materials containing chalcogenides, also rather ionic, 4 materials containing nitrogen and
5 III-V materials, less ionic, and 4 materials from group-IV elements, non-ionic, except SiC.

For oxygen-based materials, the ZPR ranges from \mbox{150} meV to \mbox{700} meV, and the gFr model captures
this very well, with less than 25\% error, with only one exception, BeO.
The chalcogenide materials are also reasonably well described by the gFr model, capturing at least two-third of the ZPR. Globally their ZPR is smaller (note the logarithmic scale).

For the nitride materials and for SiC, the gFr captures about 50\% of the quite large ZPR (between \mbox{176 meV} and \mbox{406 meV}). The adequacy of the gFr model decreases still with the III-V materials and the three non-ionic IV materials. In the latter case, the vanishing Born effective charges result in a null ZPR within the gFr model. (These three materials are omitted from Fig.~\ref{fig:FROvsFPb_gap_fourth}).

For the oxydes and chalcogenides,
{\it the ZPR is thus dominated by the zone-center parameters} (including the phonon frequencies), and the physics corresponds to the one of the large-polaron picture~\cite{Feynman1955}, namely,
{\it the slow electron motion is correlated to a phonon cloud that
dynamically adjusts to it}. This physics is completely absent from the ASC approach.
Even for nitrides, the gFr describes a significant fraction of the ZPR.

A perfect agreement between the non-adiabatic AHC first-principles ZPR and the generalized Fr\"ohlich model ZPR
is not expected. Indeed, differences can arise from different effects: lack of dominance
of the Fr\"ohlich {\color{black} electron-phonon interaction} in some regions of the Brillouin Zone, departure from parabolicity of the electronic structure
(obviously, the electronic structure must be periodic so that the parabolic behavior does not extend to infinity), interband contributions,
phonon band dispersion, incomplete cancellation between the {\color{black} Debye-Waller} and the acoustic phonon mode contribution.

It is actually surprising to see that
for so many materials, the generalized Fr\"ohlich model matches largely the first-principles AHC results.
Anyhow, as a conclusion for this section,  for a large number of materials, {\it we have validated, a posteriori and from first principles,
the relevance of large-polaron research based on Fr\"ohlich model} despite the numerous approximations on which it relies.

\section*{Discussion}
\label{sec:Discuss}

We focus on the mechanism by which the AHC divergence
{\color{black} of the ZPR}  in the adiabatic case for {\color{black} infrared}-active materials~\cite{Ponce2015} is avoided, either using the ASC methodology or using the non-adiabatic AHC methodology.
As Fr\"ohlich and Feynman have cautioned us~\cite{Froehlich1954, Feynman1955}, and already
mentioned briefly in previous sections, the dynamics of the ``slow'' electron is crucial
in this electron-phonon problem.

\begin{figure}
\includegraphics[width=0.48\textwidth]{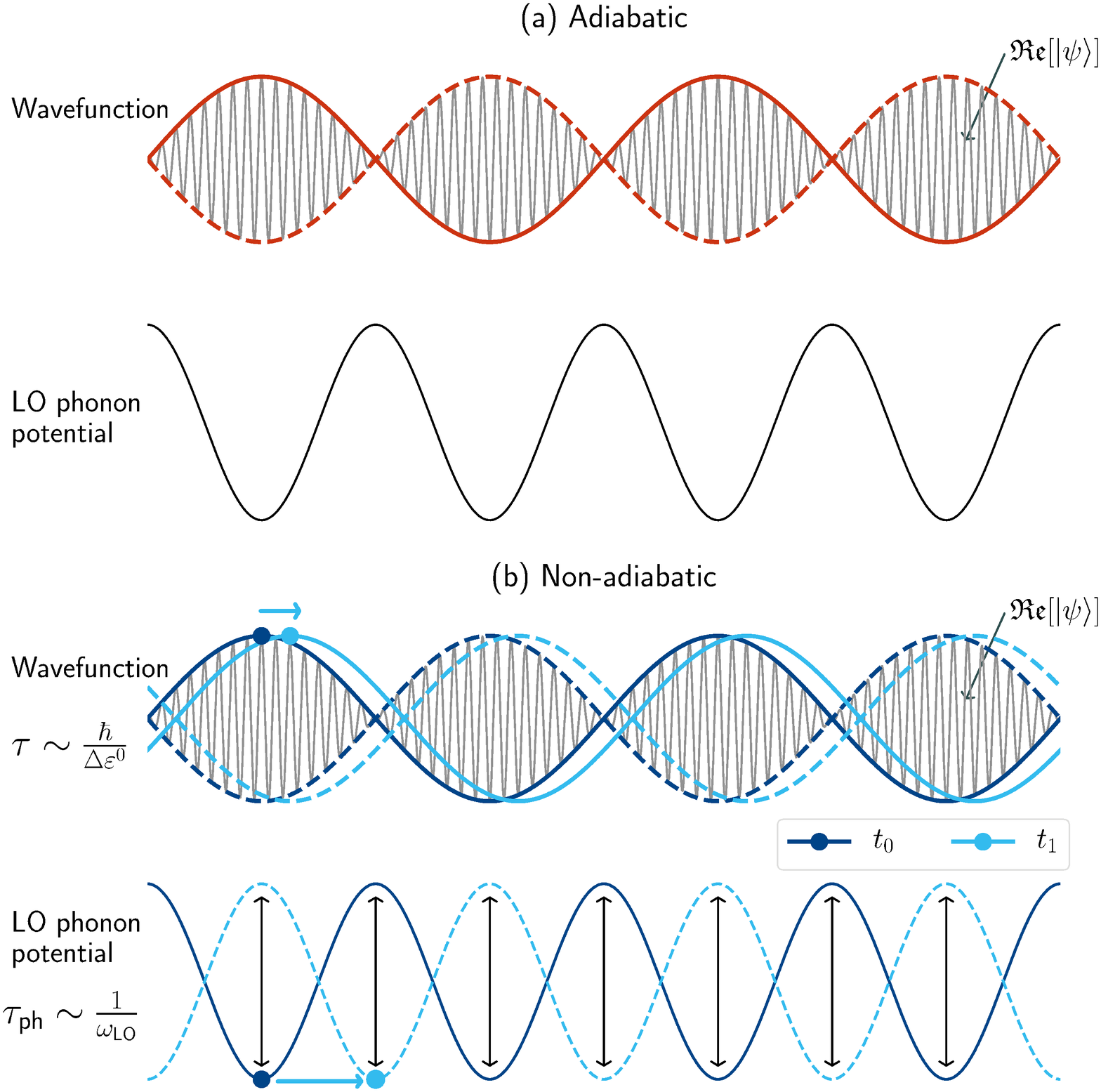}
\caption{ \label{fig:Schema}
Schematic representation of the long-wave phonon-induced potential and corresponding electronic wavefunction, real part (oscillating with lattice periodicity) and envelope:
(a) Adiabatic case, (b) non-adiabatic, time-dependent case. In the non-adiabatic case, the electron does not have the time to adjust to the change of potential, see text.}
\end{figure}

In the ASC approach, the bypass of this divergence can be understood as follows, see Fig.~\ref{fig:Schema}(a).
Consider a long-wavelength fluctuation of the atomic positions, {\it frozen in time}.
At large but finite wavelength,
the potential is periodically lowered in some regions of space and increased in some other regions of space, in an oscillatory manner with periodicity $\Delta L \propto 1/q$, where $\mathbf{q}$ is the small wavevector of the fluctuation, see Fig.~\ref{fig:Schema}(a) "LO phonon potential" part. Without such long-wavelength potential, the electron at the minimum of the conduction band has a Bloch type wavefunction, with an envelope phase factor characterized by the wavevector $\mathbf{k}_{\rm c}$ multiplying a lattice-periodic function.
Its density is lattice periodic.
With such long-wavelength potential, as a function of the amplitude of the atomic displacements,
the corresponding electronic eigenenergy changes first quadratically (as the average of the lowering and increase of potential for this Bloch wavefunction forbids a linear behaviour except in case of degeneracy), but for larger
amplitudes, it behaves linearly, as the electron localizes in the lowered potential region and the minimum of the potential
is linear in the amplitude of the atomic displacements.
This is referred to as "nonquadratic coupling" in Ref.~\onlinecite{Monserrat2015}. A wavepacket is formed, by combining Bloch wavefunctions with similar lattice periodic functions but slightly different wavevectors ($\mathbf{k}_{\rm c}$, $\mathbf{k}_{\rm c}+\mathbf{q}$, $\mathbf{k}_{\rm c}-\mathbf{q}$, etc),
coming from a small interval of energy $\Delta \epsilon^0 \propto q^2 \propto 1/(\Delta L)^2$, see Fig.~\ref{fig:Schema}(a) "Wavefunction" part.
This nonquadratic effect is actually
illustrated in Fig. 4 of Ref.~\onlinecite{Antonius2015} (see the frozen-phonon eigenvalues), as well as in Fig. 2 of Ref.~\onlinecite{Monserrat2015}.

By contrast, in the time-dependent case, as illustrated
in Fig.~\ref{fig:Schema}(b),
the wavepacket will require a characteristic time $\Delta \tau $, to form or to displace. This will be given by the Heisenberg uncertainty relation, $\Delta \epsilon^0 \Delta \tau \gtrsim \hbar$,
hence $\Delta\tau \gtrsim \hbar / \Delta\epsilon^0 \propto \Delta L^2$. For long wavelengths, the characteristic time diverges. As soon as $\Delta\tau$ is larger than the phonon characteristic time
$\tau_{\rm ph} \sim 1/\omega_{{\rm LO}}$, the ``slow'' electron will lag behind the phonon, and the static or adiabatic picture described above is no longer valid.

In all adiabatic approaches, either AHC or ASC, the electron is always supposed
to have the time to adjust to the change of potential, {\it in contradiction with the time-energy uncertainty principle}.
Furthermore, the adiabatic AHC approach only considers the quadratic region for the above-mentioned dependence of eigenvalues with respect to amplitudes of displacements.
This results in a diverging term~\cite{Ponce2015}.
At variance with the AHC case, the ASC approach samples a whole set of amplitudes, including the
onset of the asymptotic linear regime, in which case the divergence does not build up.
However, this ASC picture does not capture the real physical mechanism that prevent the divergence to occur, the impossibility
for the electron to follow the phonon dynamics, that we have highlighted above. By contrast, such physical mechanism is present both
in the non-adiabatic AHC approach and in the (generalized) Fr\"ohlich model: the ``slow" electron does not follow adiabatically (instantaneously) the atomic motion.
The divergence of the adiabatic AHC is indeed avoided in the non-adiabatic picture by taking into account the non-zero phonon frequencies.

Thus the ASC avoids the adiabatic AHC divergence for the wrong reason, which explains its poor predictive capability for the more ionic materials emphasized by Fig.~\ref{fig:FPvsExp}.
To be clear, we do not pretend the nonquadratic effects are all absent, but {\it the non-adiabatic effects have precedence}{\color{black}, at least for materials with significant infrared activity},
and the nonquadratic localization effects will be observed only if the electrons have the time to physically react.
The shortcomings of the ASC approach are further developed in the SI Sec.V, and
as a consequence of such understanding, all the results obtained for strongly {\color{black} infrared}-active materials using the adiabatic frozen-phonon supercell methodology should be questioned.

{\color{black} For non-infrared-active materials, the physical picture that we have outlined, namely the inability of slow electrons to follow the dynamics of fast phonons, is still present, but does not play such a crucial role: the electron-phonon interaction by itself does not diverge in the long-wavelength limit as compared to the infrared-active electron-phonon interaction, see Eq.~(\ref{eq:SM_ggFr}), only the denominator of the Fan self-energy diverges, which nevertheless results in an integrable ZPR~\cite{Ponce2015}.
In such case, neglecting non-adiabatic effects, as in the ASC approach, is only one among many approximations done to obtain the ZPR.}

Beyond the discovery of the predominance of non-adiabatic effects in the zero-point renormalization of the band gap for many materials,
in the present large-scale first-principles study of this effect,
we have established that electron-phonon interaction diminishes the band gap
by 5\% to 10\% for materials
containing light atoms like N or O (up to 0.7 eV for BeO),
a decrease that cannot be ignored in accurate calculations of the gap.
Our methodology, the non-adiabatic Allen-Heine-Cardona
approach, has been validated by showing that,
for nearly all materials for which experimental data exists, it
achieves quantitative agreement (within 25\%) for this property.

We have also shown that most of the discrepancies with respect to experimental data of the (arguably) best
available methodology for the first-principles band-gap computation,
denoted GWeh, originate from the
first-principles zero-point renormalization: after including it, the average overestimation from GWeh nearly vanishes.
There are some exceptions, materials in which transition metals are present, for which the addition of zero-point renormalization worsens the agreement of the band gap.
For the latter materials,
we believe that the GWeh approach is not accurate enough.

\section*{METHODS}
\label{sec:Methods}

\noindent{\bf{First-principles electronic and phonon band structures.}} Calculations have been performed using ABINIT~\cite{Gonze2016} with norm-conserving pseudopotentials and a plane-wave basis set.
Table~S1
of the SI provides calculation
parameters: plane-wave kinetic cut-off energy, electronic wavevector sampling in the BZ,
and largest phonon wavevector sampling in the BZ.
For most of the materials, the GGA-PBE exchange-correlation functional~\cite{Perdew1996} has been used and
the pseudopotentials have been taken from the PseudoDojo project~\cite{Setten2018}.
For diamond, BN-zb, and AlN-wz,
results reported here come from Ref.~\onlinecite{Ponce2015}, where the LDA has been used,
with other types of pseudopotentials.

The calculations have been performed at the theoretical optimized lattice parameter, except for Ge
for which the gap closes at such parameter, for GaP, as at such parameter the conduction band presents unphysical quasi-degenerate valleys, and for TiO$_2$, as the GGA-PBE predicted structure is unstable~\cite{Montanari2002}.
For these, we have used the experimental lattice parameter.
The case of SrTiO$_3$ is specific and will be explained later.

Density-functional perturbation theory~\cite{Baroni2001,Gonze1997,Laflamme2016}
has been used for the phonon frequencies, dielectric tensors, Born effective charges,
effective masses, and electron-phonon matrix elements.

\noindent{\bf{First-principles calculations of zero-point renormalization.}}
{\color{black} We first detail the method used for the AHC calculations}. In the many-body perturbation
theory approach, an electronic self-energy $\Sigma$ appears due to the {\color{black} electron-phonon interaction},
with Fan and {\color{black} Debye-Waller} contributions at the lowest order
of perturbation~\cite{Giustino2017}:
\begin{eqnarray}
\Sigma_{\mathbf{k}n}(\omega) = \Sigma^{\mathrm{Fan}}_{\mathbf{k}n}(\omega) + \Sigma^{\mathrm{DW}}_{\mathbf{k}n}.
\label{eq:self_energy_DW+Fan}
\end{eqnarray}
The Hartree atomic unit system is used throughout ($\hbar=m_e=e=1$).
An electronic state is characterized by $\mathbf{k}$, its wavevector, and $n$, its band index, $\omega$ being the frequency.
These two contributions correspond to the two diagrams
presented in Fig.~\ref{fig:FanDW}.

Approximating the electronic Green's function by its non-interacting KS-DFT counterpart without {\color{black} electron-phonon interaction},
gives the standard result for the $T=0$~K retarded Fan self-energy \cite{Giustino2017}:
\begin{eqnarray}
&&\Sigma^{\mathrm{Fan}}_{\mathbf{k}n}(\omega) = \frac{1}{N_{\mathbf{q}}} \sum_{\mathbf{q}j}^{\rm BZ} \sum_{n'} |\langle \mathbf{k+q}n'|H^{(1)}_{\mathbf{q}j}|\mathbf{k}n\rangle|^2 \times \nonumber \\
&&\left[
\frac{ 1-f_{\mathbf{k+q}n'}}{\omega-\varepsilon_{\mathbf{k+q}n'}-
\omega_{\mathbf{q}j}+i \eta}+
\frac{ f_{\mathbf{k+q}n'}}{\omega-\varepsilon_{\mathbf{k+q}n'}+
  \omega_{\textbf{q}j}+i \eta} \right].
\nonumber
\\
\label{eq:self_energy_Fan}
\end{eqnarray}

In this expression, contributions from phonon modes with harmonic
phonon energy $\omega_{\mathbf{q}j}$
are summed for all branches $j$, and wavevectors $\mathbf{q}$, in the entire Brillouin Zone (BZ).
The limit for infinite number $N_{\mathbf{q}}$ of wavevectors (homogeneous sampling) is implied.
Contributions from transitions to electronic states $\langle\mathbf{k+q}n'|$ with
KS-DFT electron energy $\varepsilon_{\mathbf{k+q}n'}$
and occupation number $f_{\mathbf{k+q}n'}$ (1 for valence, 0 for conduction, at $T=0$~K)
are summed for all bands $n'$ (valence and conduction).
The $H^{(1)}_{\mathbf{q}j}$ is the self-consistent change of potential due to the $\mathbf{q}j$-phonon~\cite{Giustino2017}.
Limit of this expression for vanishing positive $\eta$ is implied.
For the {\color{black} Debye-Waller} self-energy, $\Sigma^{\mathrm{DW}}_{\mathbf{k}n}$, we refer to Refs.~\onlinecite{Ponce2014a,Giustino2017}.

In the non-adiabatic AHC approach, the ZPR is obtained directly from the real part of the self-energy, Eq.~(\ref{eq:self_energy_DW+Fan}),
evaluated at $\omega=\varepsilon_{\mathbf{k}n}$~\cite{Giustino2017}:
\begin{equation}
{\rm ZPR}_{\mathbf{k}n}^{\rm AHC}=
\Re e \Sigma_{\mathbf{k}n}(\omega=\varepsilon_{\mathbf{k}n}).
\label{eq:ERS8}
\end{equation}
If the adiabatic approximation is made, the phonon frequencies $\omega_{\mathbf{q}j}$ are considered small with respect to eigenenergy differences
in the denominator of Eq.~(\ref{eq:self_energy_Fan}) and are simply dropped,
while a finite $\eta$, usually 0.1 eV, is kept.
With a vanishing $\eta$, the adiabatic AHC ZPR at band edges diverges for {\color{black} infrared}-active materials, see SI Sec.IV.A.

Summing the Fan and {\color{black} Debye-Waller} self-energies,
and working also with the  rigid-ion
approximation for the {\color{black} Debye-Waller} contribution
delivers the non-adiabatic AHC ZPR, given explicitly in Eqs.~(16) and (17) of Ref.~\onlinecite{Ponce2015}.
We do not work with the approach called
dynamical AHC, also mentioned in
Ref.~\onlinecite{Ponce2015}.
It corresponds to
Eq.~(166) of Ref.~\onlinecite{Giustino2017}.
Both non-adiabatic and dynamical AHC ZPR flavors were studied e.g. in Ref.~\onlinecite{Antonius2015},
but the comparison with the diagrammatic quantum Monte Carlo results for the Fr\"ohlich model, see e.g. Fig. 1 of Ref.~\onlinecite{Nery2018}, is clearly in favor of the
non-adiabatic AHC approach. Actually, this also constitutes a counter-argument to the claim
by Cannuccia and Marini, Ref.~\onlinecite{Cannuccia2011}, that
band theory might not apply to carbon-based nanostructures.
As shown in Ref.~\onlinecite{Nery2018}, the cumulant expansion results for the spectral function demonstrates that the dynamical AHC spectral function is unphysical, with only one wrongly placed satellite.
The physical content of the present non-adiabatic AHC theory, focusing on the crucial role of the LO phonons, is very different from the physical analysis of Ref.~\onlinecite{Cannuccia2011}, based on the dynamical AHC theory.

The imaginary smearing of the denominator
in the ZPR computation is 0.01eV, except for
SiC, where 0.001 eV is used.
Other technical details are similar to previous
studies by some of ours, e.g. Refs.~\onlinecite{Ponce2015} and \onlinecite{Nery2018}.

The dependence of the electronic structure on zero-point lattice parameter corrections is computed from
\begin{equation}
{\rm ZPR}_{\mathbf{k}n}^{\rm lat}={\rm \varepsilon}_{\mathbf{k}n}(\{{\bf{R}}_i^{T=0}\})
-
{\rm \varepsilon}_{\mathbf{k}n}(\{{\bf{R}}_i^{\rm fix}\}),
\label{eq:SM_ZPRlat}
\end{equation}
where the lattice parameters $\{{\bf{R}}_i^{\rm fix}\}$ minimize the Born-Oppenheimer energy without phonon contribution, while
$\{{\bf{R}}_i^{T=0}\}$
minimizes the free energy that includes zero-point phonon contributions, see SI Sec.II.C.

{\color{black} At variance with the AHC approach, in the ASC
the temperature-dependent average band edges (here written for the bottom of the conduction band) are obtained from
\begin{eqnarray}
\langle \varepsilon_{\rm c}(T) \rangle =Z_I ^{-1} \sum_m \exp(- \beta E_m) \langle\varepsilon_{\rm c} \rangle_m,
\label{eq:eigenASC}
\end{eqnarray}
where $\beta=k_BT$ (with $k_B$ the Boltzmann constant), $T$ the temperature, $Z_I$ the canonical partition function among the quantum nuclear states $m$ with energies $E_m$
($Z_I=\sum_m \exp(- \beta E_m)$),
and $\langle\varepsilon_{\rm c} \rangle_m$
the band edge average taken over the corresponding many-body nuclear wavefunction.
At zero Kelvin, this gives an instantaneous average of the band edge value over zero-point atomic displacements, computed while the electron is NOT present in the conduction band (or hole in the valence band)
thus suppressing all correlations between the phonons and the added (or removed) electron.}

\noindent{\bf{Convergence of the calculation.}}
As previously noted~\cite{Ponce2015,Nery2018}, the sampling of phonon wavevectors in the Brillouin zone is a delicate issue,
and has been thoroughly analyzed in Sec.IV.B.2 of
Ref.~\onlinecite{Ponce2015}. In particular,
for {\color{black} infrared}-active materials treated with the non-adiabatic effects, at the band structure extrema, a $N_{\mathbf{q}}^{-1/3}$
convergence of the value
is obtained.
We have taken advantage of the knowledge acquired in Ref.~\onlinecite{Ponce2015} to
accelerate the convergence by three different methodologies. In the first one, we fit the $N_{\mathbf{q}}^{-1/3}$ behavior
for grids of different sizes, and extrapolate to infinite $N_{\mathbf{q}}$.
In the second one, we estimate the missing contribution to the integral around $\mathbf{q}=0$,
at the lowest order, see SI Sec.I.B, using ingredients similar to
those needed for the
generalized Fr\"ohlich model, except the effective masses.
For ZnO and SrTiO$_3$, a third correcting scheme,
further refining the region around the band edge with an extremely fine grid, is used.

\noindent{\bf{The special case of SrTiO$_3$.}}
While phonons in most materials in the present study are well addressed within the harmonic approximation, this is not the case for SrTiO$_3$.
This material is found in the cubic perovskite structure at room temperature, and undergoes a transition to a tetragonal phase below $150$~K, characterized by tilting of the TiO$_6$ octahedra~\cite{GuennouPressuretemperaturephasediagram2010}.
Within our first-principles scheme in the adiabatic approximation, the cubic phase remains unstable with respect to tilting of the octahedra.
Quantum fluctuations of the atomic positions actually play a critical role in stabilizing the cubic phase at high temperature~\cite{Tadano2015},
as well as suppressing the ferroelectric phase at low temperature
~\cite{Muller1979,Wang1996}.

Since SrTiO$_3$ is however a material for which large polaron
effects are clearly identified, see e.g. Ref.~\onlinecite{Devreese2010} and references therein, we decided to study it as well.
We addressed the anharmonic stabilization problem using the state-of-the-art TDep methodology~\cite{Hellman2013}.
We used VASP molecular dynamics to generate 40 configurations in a 2x2x2 cubic cell of STO at $300$~K,
producing 20,000 steps with 2fs per step and sampling the 40 configurations out of the last 5000 steps.
Then we computed the forces with ABINIT and
{\color{black} performed TDep calculations with the ALAMODE code \cite{Tadano2014}}.
Our calculation
stabilizes the acoustic phonon branches and yields a phonon band structure in good agreement with experimental data.

\noindent{\bf{Sources of discrepancies between experiment and theory.}}
The anharmonic corrections to phonon frequencies are not the only reasons for potential differences between the
experimental ZPR$_{\rm g}$ and our non-adiabatic AHC ZPR$_{\rm g}$ values. The following phenomena may also play a role:
(1) the rigid-ion approximation~\cite{Gonze2011a,Ponce2014a};
(2) the nonquadratic behaviour of the eigenenergies with collective displacements of the nuclei, in reference to the ASC, especially emphasized in Ref.~\onlinecite{Monserrat2015};
(3) the reliance on GGA-PBE eigenenergies and eigenfunctions, instead of more accurate (e.g. GW) ones,
as in Ref.~\onlinecite{Antonius2014}, also discussed in Ref.~\onlinecite{Karsai2018};
(4) self-trapping effects, overcoming the quantum fluctuations, yielding small polarons, see e.g. Ref.~\onlinecite{Sio2019}.
There is still little knowledge about each of these effects when correctly combined to predict the ZPR$_{\rm g}$ beyond the AHC picture.

As an example, in Ref.~\onlinecite{Karsai2018}, the difference between the ASC-PBE and
the ASC-GW was argued to be only a few meV,
but a more careful look at their values show that it is often bigger than 10\% of the ASC-PBE.
Unfortunately, the convergence of the ASC-GW
results with respect to supercell size could not be convincingly achieved in
Ref.~\onlinecite{Karsai2018}.
It remains to be seen whether a non-adiabatic AHC treatment based on GW matrix elements would differ by such relative ratio, see Sec.V.C of the SI.
Altogether, it would
be hard to claim more than 25\% accuracy with respect to experimental data, from our non-adiabatic AHC ZPR$_{\rm g}$ calculations.
Together with the experimental uncertainties, this explains our choice for the 25\% accuracy comparative limit used in Fig.~\ref{fig:FPvsExp}.

\section*{ACKNOWLEDGMENTS}
We acknowledge fruitful discussions with Y. Gillet and S. Ponc\'e.
This work has been supported by the Fonds de la Recherche Scientifique (FRS-FNRS Belgium) through the PdR Grant No. T.0238.13 - AIXPHO,
the PdR Grant No. T.0103.19 - ALPS, the Fonds de Recherche du Qu\'ebec Nature et Technologie (FRQ-NT), the Natural Sciences and Engineering Research Council of Canada (NSERC) under grants RGPIN-2016-06666. Computational resources have been provided by the supercomputing facilities
of the Universit\'e catholique de Louvain (CISM/UCL), the Consortium des Equipements de Calcul Intensif en
F\'ed\'eration Wallonie Bruxelles (CECI) funded by the FRS-FNRS under Grant No. 2.5020.11,
the Tier-1 supercomputer of the F\'ed\'eration Wallonie-Bruxelles, infrastructure funded by the Walloon Region under the grant agreement No. 1117545,
as well as the Canadian Foundation for Innovation, the Minist\`ere de l'\'Education des Loisirs et du Sport (Qu\'ebec), Calcul Qu\'ebec, and Compute Canada.
This work was supported by the Center for Computational Study of Excited-State Phenomena in Energy Materials (C2SEPEM) at the Lawrence Berkeley National Laboratory, which is funded by the U.S. Department of Energy, Office of Science, Basic Energy Sciences, Materials Sciences and Engineering Division under Contract No. DE-AC02-05CH11231, as part of the Computational Materials Sciences Program. This research used resources of the National Energy Research Scientific Computing Center (NERSC), a DOE Office of Science User Facility supported by the Office of Science of the U.S. Department of Energy under Contract No. DE-AC02-05CH11231.

\section*{AUTHOR CONTRIBUTIONS}
A. Miglio has conducted calculations for most oxyde materials, with help from M. Giantomassi. V. Brousseau-Couture has conducted calculations for most other materials. G. Antonius and Yang-Hao Chan
have conducted calculations for SrTiO$_3$ and ZnO.
E. Godbout has conducted calculations of the lattice ZPR of 6 materials. X. Gonze has worked out the generalized Fr\"ohlich model and perturbative treatment.
X. Gonze and M. C\^ot\'e have supervised the work. All authors have contributed to the writing of the manuscript.

\section*{COMPETING INTERESTS}
The authors declare no competing financial or non-financial interests.

\section*{DATA AVAILABILITY}
The numerical data used to
create all the figures in the main text have been collected in the the Supplementary Information.

\section*{CODE AVAILABILITY}
ABINIT is available under GNU {\color{black} General Public Licence} from the ABINIT web site (http://www.abinit.org).

\newpage
\section*{SUPPLEMENTARY INFORMATION}

In the Supplementary Information,
the tables with the numerical data used to
create the figures in the main text have been collected,
as well as
information and comments about such data (including the
corresponding bibliographical references for results that have not been computed in the present work).
Experimental data for the ZPR are critically examined and discussed.
The gap ZPR values are also discussed with respect to previously published values, highlighting the importance of the control of phonon wavevector grid convergence and imaginary broadening.
The non-quadratic effects and GW approximation versus DFT approach are further discussed in the light of previously published data.

\section{First-principles AHC calculations: list of materials, parameters, accuracy}
\label{sec:FP}

\subsection{List of materials and calculation parameters}
\label{sec:List}

We have evaluated 
Eqs.~(8) and (9) of the main text for thirty materials.
The list of these materials, with Materials Project IDs~\cite{MP}, is presented in Table
\ref{tab:SMparam}.
It includes three non-polar
materials (C, Si, Ge), and one of their combinations (SiC),
nine III-V compounds (GaAs, AlAs, AlP, GaN-w, GaN-zb, AlN, BN, AlSb, GaP) among which four nitrides,
ten oxides (TiO$_2$, ZnO, SnO$_2$, BaO, SrO, CaO, Li$_2$O, MgO, SiO$_2$-stishovite, BeO) and SrTiO$_3$, as well as
six non-oxide II-VI compounds (CdTe, CdSe, CdS, ZnS, ZnSe, ZnTe).

With this set of materials, we have (i) a large overlap with well established experimental ZPR$_{\rm g}$ values,
as discussed in Sec.~\ref{sec:ZPR_Exp}, and complete coverage of the set of materials
for which ASC ZPR$_{\rm g}$ calculations
have been done by Karsai {\it et al.}~\cite{Karsai2018};
(ii) nearly entire coverage of the set of materials for which GWeh
computations have been performed, from Shishkin {\it et al.}~\cite{Shishkin2007a}; and we include also
(iii) a dozen of binary oxyde infrared-active materials for which the trends in ZPR can be analyzed.

Table
\ref{tab:SMparam} also lists the
parameters  used  in  the  first-principles  computations.
Within Density-Functional Perturbation Theory (see METHODS section), the computation of phonons at arbitrary q-points is done without any supercell calculation, thus allowing us to rely on the very fine q-point sampling grids mentioned in Table
\ref{tab:SMparam} at affordable CPU time cost.
Still the convergence with respect to such sampling is extremely slow,
and is adressed below.

\begin{table}[h]
\renewcommand\thetable{S1}
\caption{\label{tab:SMparam}
List of the thirty materials considered
in the present study, with parameters
used in the first-principles computations.
From left to right: material; Materials Project ID~\cite{MP}; energy cut-off for the planewave basis set;
wavevector sampling for the electronic states; wavevector sampling for the phonon states.
Abbreviations for crystal structures: sc=simple cubic, rs=rocksalt, zb=zinc-blende, w=wurtzite, dia=diamond, t=tetragonal, rh=rhombohedral.
}
\begin{tabular}{ | l | l | l | c | c |}
 \hline \hline
                 && E$_{\rm cut}$ &  {\bf k}-point   &  {\bf q}-point \\
                 &&                & sampling  & sampling \\
 Material    & ID & (Ha)          &                 &                \\
\hline \hline
Ge-dia       & mp-32     & 40  &6x6x6 (x4shifts) &48x48x48  \\
Si-dia         &mp-149   &  20 & 6x6x6 (x4shifts) & 100x100x100 \\
GaAs-zb     &mp-2534 & 40 &6x6x6 (x4shifts)&64x64x64 \\
CdTe-zb     &mp-406   & 50 &6x6x6 (x4shifts)&48x48x48 \\
AlSb-zb     &mp-2624   & 40 & 6x6x6 (x4shifts)& 48x48x48 \\
CdSe-zb     &mp-2691 & 50  &6x6x6 (x4shifts)&48x48x48 \\
AlAs-zb      &mp-2172  & 40 &6x6x6 (x4shifts)&48x48x48  \\
ZnTe-zb     &mp-2176 & 40 & 6x6x6 (x4shifts) & 48x48x48 \\
GaP-zb     &mp-2690   & 40 & 6x6x6 (x4shifts)& 48x48x48 \\
SiC-zb        &mp-8062  & 35 &6x6x6 (x4shifts)&48x48x48  \\
CdS-zb      &mp-2469  &45 &6x6x6 (x4shifts)&48x48x48 \\
AlP-zb       & mp-1550   & 25 &8x8x8 (x4shifts)&48x48x48    \\
ZnSe-zb     &mp-1190 & 40 & 6x6x6 (x4shifts) & 48x48x48 \\
TiO$_2$-t  &mp-2657 & 40 & 6x6x8 & 20x20x32 \\
SrTiO$_3$-sc &mp-5229 & 70  & 8x8x8 & 48x48x48 \\
GaN-w       &mp-804 & 40 &8x8x8 &64x64x64  \\
GaN-zb     &mp-830 & 40 &6x6x6 (x4shifts)&48x48x48 \\
ZnO-w       &mp-2133 &  50 & 6x6x6 & 48x48x48 \\
SnO$_2$-t &mp-856 & 40 & 6x6x8 & 20x20x32   \\
ZnS-zb      &mp-10695  &40  &6x6x6 (x4shifts) &48x48x48  \\
BaO-rs        &mp-1342 & 40 & 8x8x8 & 32x32x32 \\
SrO-rs        &mp-2472 & 40 & 8x8x8 & 32x32x32  \\
C-dia         &mp-66    & 30  & 6x6x6 (x4shifts) & 125x125x125 \\
AlN-w        &mp-661 &  35 & 6x6x6 & 34x34x34 \\
BN-zb        &mp-1639 & 35 & 8x8x8 (x4shifts) & 100x100x100 \\
CaO-rs       &mp-2605 & 40 & 8x8x8 & 32x32x32 \\
Li$_2$O      &mp-1960 & 50 & 8x8x8 & 32x32x32 \\
MgO-rs      &mp-1265 & 50 & 8x8x8 (x4shifts) & 96x96x96\\
SiO$_2$-t   &mp-6947 & 40 & 6x6x8 & 20x20x32 \\
BeO-w       &mp-2542 & 50 & 12x12x6 & 32x32x16 \\
\hline \hline
\end{tabular}
\end{table}

\subsection{Accuracy of the Brillouin Zone sampling in the AHC case}
\label{sec:Accuracy}

As mentioned in the METHODS section of the main text, the convergence with respect
to the {\bf q}-point sampling
can be improved by three different techniques: either by a linear extrapolation based on the expected scaling
$(N_{\mathbf{q}})^{-1/3}$, or by the computation
of the coefficient of such scaling behaviour, as explained below, or by explicitly refining the region around $\Gamma$.
For ZnO and SrTiO$_3$, we have used the latter, with a 192x192x192 fine grid.

The second methodology is the following.
According to Ref.~\onlinecite{Ponce2015}, the $(N_{\mathbf{q}})^{-1/3}$ behavior is associated
with a $q=0$ discarded (missing) part of the Brillouin zone in the Fan contribution, Eq.~(8) of the main text.
We denote $\Omega_{q=0}$ this region of the BZ, and estimate its contribution using the
generalized Fr\"ohlich electron-phonon interaction,
however neglecting the electronic dispersion in this small region
(unlike in the generalized Fr\"ohlich model
of the main text).
The estimation of the dominant correction to the ZPR$^{\rm AHC}_{\rm c}$ thus reads
\begin{eqnarray}
\Delta^{q=0}{\rm ZPR}^{\rm AHC}_{c}&=&
-\frac{1}{\pi \Omega_0}
\int\limits_{\Omega_{q=0}} d\mathbf{q}
\sum_{j}
\Big(
\frac{ \hat{\mathbf{q}}.\mathbf{p}_j(\hat{\mathbf{q}}) }
        { q \omega_{j0}(\hat{\mathbf{q}}) \epsilon^\infty(\hat{\mathbf{q}}) }
\Big)^2.
\nonumber
\\
\label{eq:SM_ZPR_corr}
\end{eqnarray}
Note that the sum over degenerate states has been cancelled by the $n_{\rm deg}$ denominator.
The volume of the missing region is actually the BZ volume ($\Omega_{BZ}=(2\pi)^3/\Omega_{0}$)
divided by $N_{\mathbf{q}}$, the number of points sampling the BZ.
The equivalent sphere has a cut-off radius
\begin{eqnarray}
q_{\rm c}=
2 \pi \Big(
\frac{3} {4 \pi \Omega_{0} }
\Big)^{1/3} (N_{\mathbf{q}})^{-1/3}.
\label{eq:SM_qc}
\end{eqnarray}
We then replace the integral over $\Omega_{q=0}$ by an integral in the sphere with radius $q_{\rm c}$,
evaluate the radial part of the integral, and obtain
\begin{eqnarray}
\Delta^{q=0}{\rm ZPR}^{\rm AHC}_{\rm c}&\approx&
-\frac{8 \pi}{\Omega_{0}}
 \Big(
\frac{3} {4 \pi  \Omega_{0} }
\Big)^{1/3}  (N_{\mathbf{q}})^{-1/3} K_{\rm av},
\label{eq:SM_ZPR_corr_approx}
\end{eqnarray}
where $K_{\rm av}$ is the angular average of the $\hat{\mathbf{q}}$-dependent quantities in Eq.~(\ref{eq:SM_ZPR_corr}), namely,
\begin{eqnarray}
K_{\rm av}&=&
\frac{1}{4\pi}
\int_{4\pi} d\hat{\mathbf{q}}
\sum_{j}
\Big(
\frac{ \hat{\mathbf{q}}.\mathbf{p}_j(\hat{\mathbf{q}}) }
        { \omega_{j0}(\hat{\mathbf{q}}) \epsilon^\infty(\hat{\mathbf{q}}) }
\Big)^2.
\label{eq:SM_Kav}
\end{eqnarray}
The correction is of opposite sign for the top of the valence band, due to the change of occupation number.

This technique and the linear extrapolation technique have been tested for most materials in our list.
Later (Tables~S2 and S5), we will report
the values obtained with the
correction from Eq.~(\ref{eq:SM_ZPR_corr_approx}), except
for the following materials: the non-infrared-active materials (C, Si and Ge),
for which there is no such correction, and also AlN-w and BN-zb, for which the convergence study had been
done in Ref.~\onlinecite{Ponce2015}, based on the linear extrapolation~\cite{note-AlN-BN-convergence}.

Fig.~S1
compares Eq.~(\ref{eq:SM_ZPR_corr_approx}) and the linear extrapolation technique in the case of MgO.
The $\mathbf{q}$-grids are ($N_{\rm MP} \times N_{\rm MP} \times N_{\rm MP}$) Monkhorst-Pack samplings of the BZ,~\cite{Monkhorst1976}
thus cubic grids shifted four times, with a total number of points $N_{\mathbf{q}}=4 \cdot N_{\rm MP}^3$.
A 10\% accuracy (20 meV in this case) is obtained from corrected points already with a $N_{\rm MP}=20$ grid
(without extrapolation), while, without correction, linear extrapolation
from the $N_{\rm MP}=20$ and $N_{\rm MP}=32$ data gives it as well. Without correction
neither extrapolation, such accuracy is only reached with the finest $N_{\rm MP}=96$ grid.
Still, the correction Eq.~(\ref{eq:SM_ZPR_corr_approx}) does not exactly
removes the $1/N_{\rm MP}$ behaviour, although it decreases it by more than a factor of ten.
The coefficient to the $(N_{\mathbf{q}})^{-1/3}$ factor computed from this equation is only approximate.

Still another technique to speed up the convergence, based on Fr\"ohlich model, but not restricted to lowest order,
has been sketched in Ref.~\onlinecite{Nery2016}.
However, in this reference, it has only fully been elaborated for materials
with isotropic characteristics.
It should be possible to develop it further using
the generalized Fr\"ohlich model introduced in the present work.


\subsection{Comparing the AHC Brillouin Zone sampling and the ASC supercell size}
\label{sec:CompareConvergence}

One might wonder why even with the above-mentioned corrections, the AHC approach needs a grid with typically $N_{\rm MP}=20$ to obtain reasonably converged values, while
for the ASC approach, results with 5x5x5 or 6x6x6 supercells, corresponding to grid $N_{\rm MP}=5$ or $N_{\rm MP}=6$ are apparently converged at the same level~\cite{Karsai2018}.
We argue that the needs of both methods are not identical, due to different physics, and the rate of convergence is quite different.

In the ASC methodology, for which the divergence of the harmonic approximation is removed by the inclusion of the nonquadratic coupling, the convergence is much faster. This is partly because, as argued in the DISCUSSION section of the main text, ``for larger amplitudes,
$\langle$the eigenenergy$\rangle$ behaves linearly, as the electron localizes in the lowered potential region and the minimum of the potential is linear in the amplitude of the atomic displacements.". This is at variance with the quadratic behaviour found from the AHC perturbative theory.

Let us examine a numerical proof of this difference. In Ref.~\onlinecite{Karsai2018}, Karsai and coauthors quote their detailed convergence study for diamond (their Table 2) from 3x3x3 to 6x6x6 supercells, with apparent convergence within 6~meV for a ZPR around 330~meV (values: {$-0.278$}, {$-0.365$}, {$-0.315$}, and {$-0.321$} eV).
With the AHC, we checked that fluctuations of ZPRs between corresponding $\mathbf{q}$-point grids are on the order of 100 meV or larger, which is in line with the results from Ref.~\onlinecite{Ponce2014b} for random wavevector grids. Fluctuations becomes less than 100 meV only above 8x8x8.
Thus, converged values are much easier to reach with the ASC methodology than with the AHC methodology.
Still, as emphasized in the main text, the ASC avoids the adiabatic AHC divergence for the wrong physical reason, and the better convergence rate of ASC is misleading, as
the converged value is not the right one.

\begin{figure}
\renewcommand\thefigure{S1}
\includegraphics[width=0.40\textwidth]{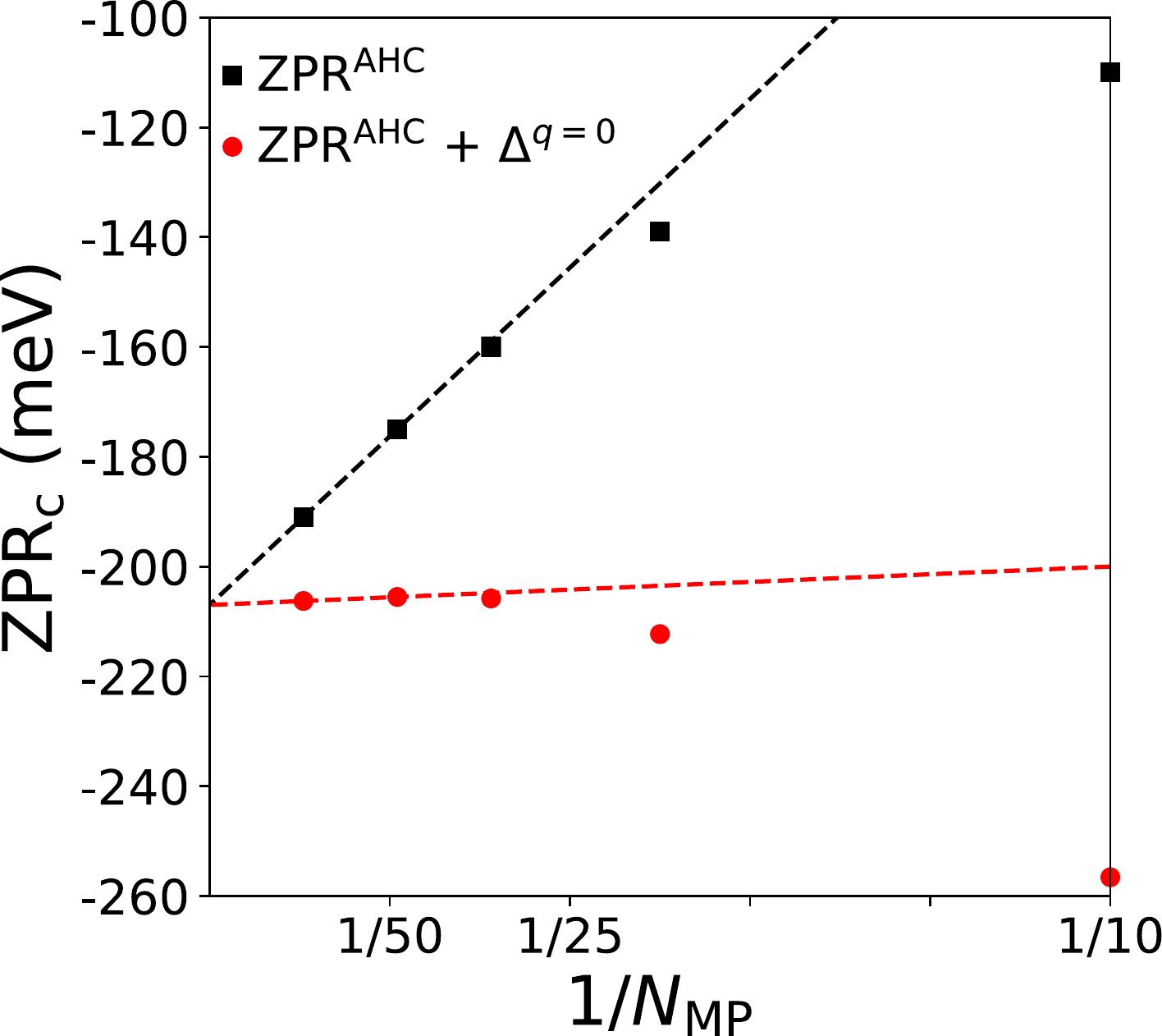}
\caption{\label{fig:ExtrapolN1q}
The ZPR$^{\rm AHC}_{\rm c}$ of MgO is computed without correction (black squares), and then with correction Eq.~(\ref{eq:SM_ZPR_corr_approx}) (red circles),
for different phonon wavevector Monkhorst-Pack $\mathbf{q}$-grid samplings,
characterized by $N_{\rm MP}$.
The points correspond to $N_{\rm MP}=96,48,32,20$ and $10$.
Dotted lines show linear extrapolation with respect to $1/N_{\rm MP}$ from the $N_{\rm MP}=48$ and $N_{\rm MP}=96$ grids, to the infinitely large $N_{\rm MP}$ value.
}
\end{figure}


\bigskip
\section{Band gap zero-point renormalization from experiment and from the theoretical approaches (AHC and ASC)}
\label{sec:ZPR_Exp_Th}

Table S2
compares ZPR$_{\rm g}$ data from experiment and from different computations.
These data are used in Fig. 2 of the main text.
Further computed ZPR$_{\rm g}$ data are provided in Table S3,
although the latter focuses on band gaps, to be analyzed in the next section (Sec.~\ref{sec:Gap}).
Gathering experimental ZPR$_{\rm g}$ data needed some care, as explained below. Also, theoretical data ought to be discussed in view of previous works.
For the in-depth discussion of the theoretical
data among themselves, we refer to Sec.~\ref{sec:ZPR_ASC}.
We also complete the list of references in which
ZPR$_{\rm g}$ have been computed, that had been mentioned in the main text.

%
\begin{table}[h]
\renewcommand\thetable{S2}
\caption{\label{tab:FPvsExp}
For eighteen materials: experimental fundamental gap and type
(d=direct, i=indirect);
experimental zero-point renormalization of the gap;
lattice contribution to the zero-point renormalization of the gap;
first-principles AHC zero-point renormalization of the gap;
 ratio $R$ between sum of AHC and lattice values and experimental values (see text) (ZPR$^{\rm AHC}_{\rm g}$+ZPR$^{\rm lat}_{\rm g}$)/ZPR$^{\rm exp}_{\rm g}$;
first-principles ASC zero-point renormalization of the gap.
Experimental ZPR$^{\rm exp}_{\rm g}$ data are obtained from
extrapolation of temperature dependence of the gap, or isotopic mass derivatives of the gap. The latter are indicated with an asterisk.
 For SiC, the experimental data refer to the 15R polymorph, while the theoretical data refer to the 3C (zb) polymorph. They have similar sp$ ^3$ topology, but differ by stacking sequence.
 For SrTiO$_3$, the ZPR$^{\rm lat}_{\rm g}$ was hardly computable with the TDep methodology presented in the METHODS section of the main text. We have estimated it combining
 experimental data for the zero-point correction to the volume~\cite{Lytle1964} and theoretical data for the derivative of the gap with volume~\cite{Zacharias2020}, see text.
 The reference experimental value for AlN is the average of the reported values from Refs.~\onlinecite{Passler2002} and~\onlinecite{Passler2003}, see text.
}
\begin{tabular}{ | l | l | l | r | r | r | r| 
}
 \hline \hline
                 & $E_{\rm g} $ & ZPR$_{\rm g}$ & ZPR$_{\rm g}^{\rm lat}$ & ZPR$_{\rm g}$ & $R$ & ZPR$_{\rm g}$ 
\\
& exp & exp & & AHC & & ASC  
\\
&       & &        &  present       & &  PBE 
\\
& [\onlinecite{note-experimentalgap}]  & & & work & & [\onlinecite{Karsai2018}] 
\\
 Material   & (eV) & (meV) & (meV) & (meV) & & (meV) 
\\
 \hline \hline
Ge-dia & 0.74 (i) & -52 [\onlinecite{Parks1994}]* &
-10
&-33 &0.83 & -50 
\\
 & & -52 [\onlinecite{Passler2002}]
  & & & &  
  \\
Si-dia & 1.17 (i) & -59 [\onlinecite{Karaiskaj2002}]* & +9 & -56 & 0.80 & -65
\\
& & -72 [\onlinecite{Passler2002}]& & & &  
\\
GaAs-zb & 1.52 (d)& -60 [\onlinecite{Passler2002}] & -29 & -18 & 0.78 & -53 
\\
CdTe-zb & 1.61 (d) & -16 [\onlinecite{Passler2002}] & -8 & -20 & 1/0.57 & -15 
\\
AlSb-zb & 1.69 (i) & -35 [\onlinecite{Passler2002}] & +6 & -51 & 0.78 & -43 
\\
CdSe-zb & 1.85 (d) & -38 [\onlinecite{Passler2002}]& -7 & -34 & 1/0.93 & -21 
\\
AlAs-zb & 2.23 (i) & -50 [\onlinecite{Passler2002}] & +8 & -74 & 1/0.76 &-63 
\\
GaP-zb & 2.34 (i)  & -85 [\onlinecite{Passler2002}] & +8 & -65 & 0.67 & -57 
\\
ZnTe-zb & 2.39 (d) & -40 [\onlinecite{Passler2002}] & -18 & -22 & 1.00 & -24 
\\
SiC & 2.40 (i) & -215 [\onlinecite{Passler2002}] & +6 & -179 & 0.80 & -109 
\\
CdS-zb & 2.58 (d) & -62 [\onlinecite{Zhang1998}]* & -10 & -70 & 1/0.78 & -29 
\\
& & -34 [\onlinecite{Passler2002}]& & & &  
\\
ZnSe-zb & 2.82 (d)& -55 [\onlinecite{Passler2002}] & -17 &-44 &1/0.90 & -28 
\\
ZnS-zb & 3.84 (d) & -105 [\onlinecite{Manjon2005}]* &-24 & -88 & 1/0.94 & -44 
\\
& & -78 [\onlinecite{Passler2002}]& & & &  
\\
C-dia & 5.48 (i) & -338 [\onlinecite{Collins1990}]* & -27 & -330 & 1/0.95 & -320 
\\
& & -334 [\onlinecite{Passler2002}] & & & &  
\\
SrTiO$_3$ & 3.25 (i) [\onlinecite{Benrekia2012}] & -336 [\onlinecite{Kok2015}] &$\approx$ +20&-290 & 0.80 & 
\\
ZnO-w & 3.44 (d) & -164 [\onlinecite{Manjon2003}]* & -17 &-157 & 1/0.94 & -57 
\\
GaN-w & 3.47 (d) & -180 [\onlinecite{Passler2002}] & -49 &-189 & 1/0.75 & -94 
\\
AlN-w & 6.20 (d) & -417 & -85& -399 & 1/0.88 & 
\\
& & -350 [\onlinecite{Passler2002}]  & & & & 
\\
& & -483 [\onlinecite{Passler2003}]  & & & & 
\\
\hline \hline
\end{tabular}
\end{table}


\begin{table}[h]
\renewcommand\thetable{S3}
\caption{\label{tab:GapExpFP}
Electronic gaps and
corrections due to the zero-point motion (ZPR$^{\rm AHC}_{\rm g}$ and ZPR$^{\rm lat}_{\rm g}$), for thirteen materials:
one-shot GW calculation of E$_{\rm g}$, without ZPR (E$^{\rm G_0W_0}_{\rm g}$);
self-consistent GW with electron-hole corrections ($E^{\rm GWeh}_{\rm g}$) which is the
best theoretical $E_{\rm g}$ without ZPR ;
first-principles ZPR$^{\rm AHC}_{\rm g}$ and
firs-principles ZPR$^{\rm lat}_{\rm g}$ (this work);
ZPR-corrected $E^{\rm GWeh}_{\rm g}$ ($E^{\rm +ZPR}_{\rm g}$),
namely $E^{\rm +ZPR}_{\rm g}$=$E^{\rm GWeh}_{\rm g}$+ZPR$^{\rm AHC}_{\rm g}$+ZPR$^{\rm lat}_{\rm g}$ ;
experimental gap ($E^{\rm exp}_{\rm g}$).
The ratios between $E^{\rm G_0W_0}_{\rm g}$ and $E^{\rm exp}_{\rm g}$, between $E^{\rm GWeh}_{\rm g}$ and $E^{\rm exp}_{\rm g}$, between  $E^{\rm GWeh}_{\rm g}$+ZPR$^{\rm AHC}_{\rm g}$ and $E^{\rm exp}_{\rm g}$
and between  $E^{\rm +ZPR}_{\rm g}$ and $E^{\rm exp}_{\rm g}$, are shown in Fig. 3,
but not reported here.
Then, the ratio $R^{\rm ZPR}_{\rm GWeh}$ between ZPR$_{\rm g}^{\rm AHC+lat}$ and
the correction brought by the scGWeh approximation $E^{\rm GWeh}_{\rm g}$ with respect to $E^{\rm G_0W_0}_{\rm g}$,
is mentioned.
The values for $E^{\rm G_0W_0}_{\rm g}$ are taken from Ref.~\onlinecite{Shishkin2007a}, except for Ge-dia, that comes from Ref.~\onlinecite{Setten2017}.
}
\begin{tabular}{ | l || r | r | r | r | r | r || r |}
 \hline \hline
                & $E^{\rm G_0W_0}_{\rm g}$ & $E^{\rm GWeh}_{\rm g}$ & ZPR$^{\rm AHC}_{\rm g}$ & ZPR$^{\rm lat}_{\rm g}$ & $E^{\rm +ZPR}_{\rm g}$ & $E^{\rm exp}_{\rm g}$ & $R^{\rm ZPR}_{\rm GWeh}$ \\
                & [\onlinecite{Shishkin2007a},\onlinecite{Setten2017}] & [\onlinecite{Shishkin2007}]  &              &            & [\onlinecite{note-experimentalgap}] & & \\
Material   &  eV  & eV  &  eV  & eV  & eV & eV  & \\
 \hline \hline
Ge-dia      & 0.53 & 0.81 &  -0.033 & -0.010 & 0.77 & 0.74 & -0.15 \\
Si-dia      & 1.12  & 1.24 & -0.056 & 0.009 &  1.19 & 1.17 & -0.39 \\
GaAs-zb     & 1.30  &1.62 &  -0.024 & -0.029 & 1.57 & 1.52 & -0.17 \\
SiC-zb      & 2.27  & 2.53 & -0.179  & 0.006 & 2.36  & 2.40 & -0.67 \\
CdS-zb      & 2.06 & 2.39  & -0.070  & -0.010 & 2.31  & 2.42 & -0.24 \\
AlP-zb      & 2.44 &  2.57 &  -0.093 & 0.010 &  2.49 & 2.45 & -0.64 \\
GaN-w       & 2.80 & 3.27 & -0.189 & -0.049 & 3.03 & 3.20 & -0.51 \\
ZnO-w       & 2.12 & 3.20 & -0.157 & -0.017 &3.03 & 3.44 & -0.16 \\
ZnS-zb      & 3.29 & 3.60 & -0.088  & -0.024 & 3.49 & 3.91 & -0.36 \\
C-dia       & 5.50 & 5.79  & -0.330 & -0.027 & 5.43  & 5.48 & -1.23 \\
BN-zb       & 6.10 & 6.59  & -0.406 & -0.017& 6.17 & 6.25 & -0.86 \\
MgO-rs      & 7.25 & 8.12  & -0.524 & -0.117 & 7.48  & 7.67 & -0.74 \\
\hline \hline
\end{tabular}
\end{table}

\subsection{Experimental data: discussion}
\label{sec:ZPR_Exp}

As explained by Cardona and Thewalt, in relation to the Table 3
of their review Ref.~\onlinecite{Cardona2005}, the experimental ZPR are obtained by two different techniques.
The first one starts
from measurements of the band gap energy for different
samples of a material with varying isotopic content.
The derivative(s) of the band gap with respect to the mass(es) are extracted, followed by extrapolation to infinite mass(es).
The second one relies on
measurements of the band gap energy  in a large temperature range, including the low-temperature regime, but also the high-temperature regime, where a linear asymptote must be reached.
In this case, the ZPR is the difference between the measured zero-temperature limit and the linear extrapolation of the high-temperature asymptote to zero temperature, as shown for Ge in Fig. 21 of Ref.~\onlinecite{Cardona2005}.
When both techniques agree, the experimental result can be reasonably trusted.

For the latter technique (extrapolation of temperature-dependent data), we rely on the
analysis of P\"assler\cite{Passler2002,Passler2003}, that
superceeds his earlier work\cite{Passler1999}
on which Ref.~\onlinecite{Cardona2005} relied.
The two publications by P\"assler\cite{Passler2002,Passler2003} yield essentially
equivalent results, except for AlN.
For this material, we have taken the mean of the two reported values as reference.
However, also in the latter case, our results are within the 25\% limit.
We have surveyed the literature after 2003, but were unable to find more reliable data.

For the isotopic extrapolation, to our surprise, a large fraction of the data provided in the related
column of Table 3 of
Ref.~\onlinecite{Cardona2005} do NOT report actual
isotopic extrapolation analysis.
We have thus extensively examined the primary literature,
and have obtained that such isotopic extrapolation
is available reliably for the following materials in our list: Ge, Si, C, CdS, ZnO and ZnS.
They are clearly identified as such in table S2.
Moreover, we have also realized that a non-negligible number of data available through
other secondary references are misleading.
We will not elaborate further on this problem, but generally speaking, we warn the reader about the unreliability of several publications in the field.
The above-mentioned works of P\"assler are apparently
free of such problem.

Anyhow, the two techniques agree very well for Ge or C, or reasonably (within 25\%) for Si and ZnS, but clearly differ for CdS (about a factor of 2). P\"assler\cite{Passler2002} does not provide data for ZnO.
In Fig.2 of the main text, we have taken the isotopic extrapolation result as reference, since the extrapolation of
temperature-dependent data is often plagued with
insufficient sampling of the high-temperature
region, which is crucial for the correct extrapolation,
as discussed at length in Ref.~\onlinecite{Passler1999}.
For the above-mentioned CdS discrepancy, the global
expected trend due to the change of mass,
comparing with CdSe and CdTe, or with the Zn series,
appears more reasonable from the isotopic technique than from extrapolation of temperature-dependent data.
Also, the comparison of the
different analyses by P\"assler, in 1999, 2002,
and 2003, Refs.~\onlinecite{Passler1999,Passler2002,Passler2003}, shows stability for a few materials, but, more generally, non-negligible variations,
at the 10-25\% level.
On this basis, one can infer that, globally speaking, the reliability
of the experimental data is on the order
of 25\%, which is the reason of our use of this value
in the main text.

Note that Karsai et al.~\cite{Karsai2018}
quotes experimental data (most from secondary sources) for about half the materials they compute, but miss many available data, especially those that do not compare well with their results.

\subsection{Theoretical AHC data: discussion}
\label{sec:ZPR_Th}

The values quoted in Tables S2 and S3 for the band gap ZPR of C, BN, and MgO, respectively \mbox{-330 meV}, \mbox{-406 meV}, and \mbox{-524 meV}, differ from those previously published in Ref.~\onlinecite{Antonius2015}, authored by some of us, giving respectively \mbox{-366 meV}, \mbox{-370 meV},
and \mbox{-341 meV}.
We emphasize that the same first-principles method
as in our current manuscript (non-adiabatic AHC) was used to obtain these data.
The differences with the present data are entirely due to the sampling of the Brillouin Zone and associated imaginary broadening factor.

In Ref.~\onlinecite{Antonius2015}, the sampling of the Brillouin Zone was already emphasized as an important technical issue, especially considering the numerical noise.
For the purpose of computing the self-energy (which was the focus of Sec.~II of Ref.~\onlinecite{Antonius2015}), or computing the importance of “anharmonic effects” in the adiabatic approximation (Sec.~III of Ref.~\onlinecite{Antonius2015}), such numerical noise was addressed by keeping a large finite imaginary broadening (between 0.1-0.4 eV) and a fixed wavevector sampling 32x32x32.
The choice of the value of the imaginary broadening was detailed in the Appendix of this paper, which was explicitly mentioned in the caption of this Table 1: ``See Appendix for the values of the $\eta$ parameters used".
Such imaginary broadening was determined to ensure removal of the noise in the self-energy, in order to produce Fig.~1 and 2, but however did not deliver an accurate absolute value for the ZPR in Table~I of Sec.~II.

Indeed, in another publication of some of ours (even slightly before Ref.~\onlinecite{Antonius2015}), namely Ref.~\onlinecite{Ponce2015}, the proper limit of vanishing imaginary broadening with much improved wavevector sampling was performed, for C and BN (up to 125x125x125 grid).
The same values for C and BN  as mentioned in our current manuscript, namely -330 meV and -406 meV, were found (with one more digit, see Table VII, column ``Non-adiabatic").
MgO was not examined in Ref.~\onlinecite{Ponce2015}, but the MgO band gap ZPR was then mentioned in Ref.~\onlinecite{Nery2018}, still another publication from some of ours, page 11, end of second paragraph of first column, 526 meV.
Such value is very similar to the values in our manuscript.
The extrapolation scheme, differing from the one chosen in the present paper, explains the difference of 2 meV for MgO.
The analysis of LiF is not done in the current work because an estimate of the Fr\"ohlich parameter for the valence bands yields $\alpha$ bigger than 8, see Ref.~\onlinecite{Nery2018},
which is beyond the validity of the present perturbative treatment.
By the way, the top of the valence band does not occur at $\Gamma$ (see e.g. Ref.~\onlinecite{MP}, which was not mentioned in Ref.~\onlinecite{Nery2018}).

\subsection{Modification of the ZPR due to lattice parameter changes}
\label{sec:ZPR_Vol}


The ZPR$^{\rm AHC}_{\rm g}$ is computed at the fixed theoretical lattice parameters and internal coordinates minimizing the
DFT energy (or fixed experimental lattice parameters in the case of Ge, GaAs and TiO$_2$).
The phonon population effects on the lattice parameters, and induced change of band gap, are not taken into account in such procedure.
This corresponds strictly to the harmonic approximation.

In the so-called quasi-harmonic approximation,
the equilibrium lattice parameters and possibly internal coordinates are relaxed in order to minimize the free energy of the system.
This makes such parameters depend on the temperature (thermal expansion), but also induces their zero-point
change at $T=0$~K. In turn, such changes modifies the gap, an effect to be accounted for
in order to compare with the experimental ZPR$_{\rm g}$.
This effect has been investigated by Garro and coworkers, Ref.~\onlinecite{Garro1996}.
It is also included in several first-principles
calculations of the ZPR, see e.g. Ref.~\onlinecite{Villegas2016}.

In the present work, we have indeed
added this effect beyond the harmonic approximation: we have computed the equilibrium lattice parameters with and without zero-point motion, then evaluated the change of gap.
In the first case, we have minimized the Born-Oppenheimer energy without phonon contribution, giving $\{{\bf{R}}_i^{fix}\}$.
In the second case, the free energy, including zero-point phonon contributions computed at different
lattice parameters and interpolated between them, was minimized, giving $\{{\bf{R}}_i^{T=0}\}$. This procedure is similar to the one described in Ref.~\onlinecite{Rignanese1996}, albeit extended to two dimensions (uniaxial and basal lattice parameters) in the case of wurtzite materials.
The differences in band gap values for such lattice parameters form the lattice ZPR modification of the electronic structure:
\begin{equation}
{\rm ZPR}_{\mathbf{k}n}^{\rm lat}={\rm \varepsilon}_{\mathbf{k}n}(\{{\bf{R}}_i^{T=0}\})
-
{\rm \varepsilon}_{\mathbf{k}n}(\{{\bf{R}}_i^{fix}\}).
\label{eq:SM_ZPRlat}
\end{equation}

In such approach, one supposes that there is no cross-influence of the modification of lattice parameters on the AHC ZPR.
Also, one relies on the approximation that the internal parameters, if not determined by symmetry,  are those that minimize the Born-Oppenheimer energy at the corresponding ${\bf{R}}_i$ lattice parameters.
In other words, the deviatoric thermal forces and their effect on the band gap are not included, but the deviatoric thermal stresses are taken into account, following Ref.~\onlinecite{Carrier2007a}.
In the list of materials of Tables S2 and S3, deviatoric thermal forces only appear for wurtzite materials, that have a quite symmetric tetrahedral environment, and thus, very small expected effect.

\subsection{A survey of zero-point renormalization computations}
\label{sec:ZPR_Refs}

In complement to the references given in the introduction of the main text, we list additional ones in view of further comparisons, without pretending to be exhaustive. Refs.
~\onlinecite{Marini2008,Giustino2010,
Cannuccia2011,Gonze2011a,Cannuccia2012,Kawai2014,Ponce2014a,Ponce2014b,
Antonius2015,Friedrich2015,Ponce2015,Villegas2016,Nery2016,Saidi2016,Molina-Sanchez2016a,Antonius2016, Villegas2017, Nery2018,
Ziaei2017, Tutchton2018, Cao2019, Querales-Flores2019, Querales-Flores2020, Lihm2020}
report calculations of the ZPR$_{\rm g}$ based on the
AHC approach with KS-DFT wavefunctions and eigenenergies.
Refs.~\onlinecite{
Capaz2005,Gonze2011a,
Han2013a,
Patrick2013,Monserrat2013,Antonius2014,Monserrat2014a,Monserrat2014b,Monserrat2015,Patrick2015a,Zacharias2015,Engel2016,Monserrat2016a,Zacharias2016,Saidi2016, Karsai2018, Monserrat2018, Bravic2019}
report calculations of the ZPR$_{\rm g}$ based on the
ASC approach, usually based on KS-DFT, but also sometimes based on GW, see Refs.~\onlinecite{
Antonius2014,Monserrat2016a,Karsai2018}. As analyzed by the present generalized Fr\"ohlich model, the ASC might be reasonably predictive for purely covalent materials (e.g. C, Si, Ge), or weakly infrared-active ones (e.g. GaAs),
however, when
the ZPR$^{\rm gFr}$ is a significant fraction of the ZPR,
the predictive capability of the ASC approaches should be questioned. Even in the covalent case, non-adiabatic
corrections can be sizeable (5...15\%), see Table VII of Ref.~\onlinecite{Ponce2015}.
The ASC GW calculations performed until now~\cite{
Antonius2014,Monserrat2016a,Karsai2018}
rely at most on 4x4x4 supercells, which is at the limit of convergence.
Going from 4x4x4 to 5x5x5 supercells for KS-DFT
indeed still brings some modification, as shown by Ref.~\onlinecite{Karsai2018}.
For this reason, the effect of relying on GW corrections
can hardly be clarified in the present context,
which focuses on the non-adiabatic effects.

In a very recent publication~\cite{Engel2020}, Engel and coworkers have proposed a new method to compute the
ZPR, combining supercell calculations with an interpolation method, still in the adiabatic approximation.
They applied it for 9 materials, a subset of those examined by Karsai and coworkers~\cite{Karsai2018}
(focusing only on covalent or weakly ionic materials).
The results that they obtain do not change the overall assessment of the adiabatic
approach that we present here:
for some materials, the agreement with experiment is slightly improved or deteriorated.

Studies of strongly infrared-active materials using the ASC include e.g. hydrogen-containing materials HF, H$_2$O, NH$_3$ and CH$_4$ in Ref.~\onlinecite{Monserrat2015}, where a giant electron-phonon
interaction in molecular crystal and related important nonquadratic coupling had been investigated, the study of
ice in Ref.~\onlinecite{Engel2016},
and the study of LiF, MgO and TiO$_2$ in Ref.~\onlinecite{Monserrat2016a}, and some of the results from Ref.~\onlinecite{Karsai2018} for IR-active materials.

In addition to these references, the electron-phonon interaction has been directly incorporated in GW calculations in Refs.~\onlinecite{Botti2013a,Lambrecht2017,Bhandari2018}.
These are further commented upon in the next section.

Table~\ref{tab:ASC_Misc} presents KS-DFT ASC data
that were not reported in Table S2, as the latter only mentioned such data from Ref.~\onlinecite{Karsai2018}.
This table~\ref{tab:ASC_Misc} still focuses on our set of materials, for which AHC data are available.
For the indirect gap of Si and of SiC, as well as both the direct and indirect gaps of C, the different ASC calculations agree very well.
The agreement is less good for GaAs and the direct gap of Si.
Anyhow, these data confirm that AHC and ASC calculations of the fundamental band gap reasonably agree
for covalent materials, but strongly disagree for strongly IR-active materials (SiC-zb, TiO$_2$-t, MgO-rs).

\begin{table}[h]
\renewcommand\thetable{S4}
\caption{\label{tab:ASC_Misc}
Experimental fundamental gap E$_{\rm g}^{\rm exp}$ and type
(d=direct, i=indirect),
first-principles zero-point renormalization of the gap ZPR$^{\rm AHC}_{\rm g}$:
from the AHC (present work), and from the ASC.
}
\begin{tabular}{ | l | l | r | r| 
}
 \hline \hline
                 & E$_{\rm g} $ & ZPR$_{\rm g}$ & ZPR$_{\rm g}$ 
                 \\
& exp & AHC & ASC  
\\
&               &  present       &  KS-DFT 
\\
& [\onlinecite{note-experimentalgap}]  & work &  
\\
 Material   & (eV) & (meV) & (meV) 
 \\
 \hline \hline
Si-dia & 1.17 (i) &-56 &
-60 [\onlinecite{Monserrat2014b}] 
\\
   & & & -57 [\onlinecite{Zacharias2016}] 
   \\
  & & &  -65 [\onlinecite{Karsai2018}]
  \\
 &  &  & -75 [\onlinecite{Zhang2020}]
  \\
& 3.40 (d) &-42 & $\approx$ -28 [\onlinecite{Monserrat2016a}]
 \\
    & & & -44 [\onlinecite{Zacharias2016}] 
    \\
GaAs-zb & 1.52 (d)&-18 & -23 [\onlinecite{Antonius2014}]
\\
     & & &  -32 [\onlinecite{Zacharias2016}]
     \\
   & & & -53 [\onlinecite{Karsai2018}] 
   \\
SiC & 2.40 (i) &-179 & -109 [\onlinecite{Monserrat2014b}]
\\
 &  &  & -109 [\onlinecite{Karsai2018}]
\\
 &  &  & -145 [\onlinecite{Zhang2020}]
\\
TiO$_2$-t & 3.03 (d) &-337 & $\approx$ -150 [\onlinecite{Monserrat2016a}]
\\
C-dia & 5.48 (i) & -330 &
 -334 [\onlinecite{Monserrat2014b}]
 \\
& & &  -345 [\onlinecite{Zacharias2016}]   
\\
  & & & -320 [\onlinecite{Karsai2018}] 
  \\
 &  &  & -437 [\onlinecite{Zhang2020}]
  \\
 & 7.07 (d) & -416 & -437 [\onlinecite{Antonius2014}]
 \\
   & & & $\approx$ -410 [\onlinecite{Monserrat2016a}] 
   \\
   & & &  -450 [\onlinecite{Zacharias2016}]   
   \\
MgO-rs & 7.67 (i) &-524 & $\approx$ -220 [\onlinecite{Monserrat2016a}]
\\
 &  &  & -281 [\onlinecite{Zhang2020}]
\\
\hline \hline
\end{tabular}
\end{table}


\section{Gap values from experiment and different computations}
\label{sec:Gap}

Table S3
compares gap values from experiment and from different computations.
These data are used in Fig. 3.
The set of 12 materials  in this table includes all those for which self-consistent GW calculations with electron-hole corrections
have been performed in Ref.~\onlinecite{Shishkin2007}, except Ne, LiF and Ar.
We did not include them in our study for the
following reasons.
Ne and Ar are strongly
affected by weak van der Waals interactions.
Much more than the other materials, they will be modified by the zero-point lattice corrections, discussed earlier.
For LiF, as mentioned previously, the estimated value of $\alpha$ show that the present perturbative treatment is likely invalid.
For all three materials, excitonic effects
are strong, and should likely be included
for meaningful comparison with experimental data.

Suppplementary Table III quotes E$^{\rm G_0W_0}_{\rm g}$ values from Ref.~\onlinecite{Shishkin2007a}, except for Ge-dia, that comes from Ref.~\onlinecite{Setten2017}.
The latter reference also presents E$^{\rm G_0W_0}_{\rm g}$ values for
many materials, including more than half of those reported in Table S3.
Many other E$^{\rm G_0W_0}_{\rm g}$ values for those materials have been published.
The spread in such values for one material can be as large
as 0.5 eV.
A comparative study of published E$^{\rm G_0W_0}_{\rm g}$ values is out-of-scope of the present publication, but is presented in Ref.~\onlinecite{Rangel2020}.
The values E$^{\rm G_0W_0}_{\rm g}$ in Table S3 have the goal to show
the typical size of the E$^{\rm G_0W_0}_{\rm g}$ to
E$^{\rm GWeh}_{\rm g}$ correction, and point out that the ZPR
correction is of the same order of magnitude for materials
with light nuclei.

In addition to the AHC and ASC methodologies, there has also been attempts to incorporate
directly the electron-phonon interaction inside the GW approach~\cite{Botti2013a,Lambrecht2017,Bhandari2018}
to obtain modified gap values or ZPR$_{\rm g}$. The first publication on the subject, by Botti and Marques~\cite{Botti2013a}
was later shown by Lambrecht, Bhandari and van Schilfgaarde~\cite{Lambrecht2017} to
ignore the range of the Fr\"ohlich interaction, yielding an erroneous integration over the Brillouin Zone. Even in the latter work, there is a factor of 2 difference for the Fr\"ohlich $\alpha$
compared to the usual definition, used in the present work.
Their ZPR$_{\rm g}$ results (Table I of Ref.~\onlinecite{Lambrecht2017} and~\onlinecite{Bhandari2018}) for MgO, GaN-zb
and SrTiO$_3$, resp. (\mbox{-219 meV}, \mbox{-67 meV}, \mbox{-404 meV})
can be compared with ours (\mbox{-517 meV}, \mbox{-171 meV}, \mbox{-290 meV}), see Table~S5. They very clearly differ,
with strong underestimation with respect to ours for MgO, GaN-zb,
and strong overestimation for SrTiO$_3$. For
SrTiO$_3$, our results agree better with the experimental
data (\mbox{-336 meV}), see Table~S2.
For MgO and GaN-zb, a direct measurement
of ZPR$_{\rm g}$ is not available. Still, experimental value for the other polymorph
of GaN, namely GaN-w (\mbox{-180 meV}) is reported in Table~S2 and matches very well our
ZPR$^{\rm AHC}_{\rm g}$ value (\mbox{-189 meV}) for GaN-w.


\section{From the adiabatic AHC approach to the generalized Fr\"ohlich Hamiltonian}
\label{sec:AHC_gFr}


In order to help understand why
the generalized Fr\"ohlich Hamiltonian
captures correctly the physics of the
Fan diagram, at variance with the
adiabatic approximation,
we summarize first the discussion of the AHC case given in Ref.~\onlinecite{Ponce2015}.
Then, we derive the generalized Fr\"ohlich Hamiltonian from Eq. (8) of the main text.

\subsection{The divergence of the adiabatic AHC for IR-active materials}
\label{sec:divergence}

Under the adiabatic approximation, the phonon frequencies $\omega_{\textbf{q}j}$ are neglected
when compared to differences between electronic eigenenergies.  In other words, one supposes that the electron (or hole) responds instantaneously to the
atomic vibrations, even for those electronic transitions with a characteristic time scale (inversely proportional
to the eigenenergy difference) larger than the phonon oscillation period. This simplification
is of course questionable for intraband transitions with small momentum transfer,
as shown in Ref.~\onlinecite{Ponce2015}.
Suppressing $\omega_{\textbf{q}j}$ in the denominators of
Eq.~(8)
, and evaluating the AHC ZPR according to Eqs.~(7) and (9)
yields
\begin{eqnarray}
{\rm ZPR}^{\rm adiab}_{\mathbf{k}n}=
&&
\Re e \frac{1}{N_{\mathbf{q}}}
\sum_{\mathbf{q}j}^{\rm BZ} \sum_{n'}
\frac{|\langle \mathbf{k+q}n|H^{(1)}_{\mathbf{q}j}|\mathbf{k}n'\rangle|^2}
{\varepsilon_{\mathbf{k}n}-\varepsilon_{\mathbf{k+q}n'}+i \eta} + \Sigma^{\mathrm{DW}}_{\mathbf{k}n}.
\nonumber
\\
\label{eq:ZPRadiabatic}
\end{eqnarray}

For band edges, in particular the top of the valence band or the bottom of the
conduction band, and considering at present non-degenerate bands, the intraband ($n^\prime =n$) contribution when $\eta=0$ has two different types
of divergences for $q \rightarrow 0$, reinforcing each others:
(i) for IR-active materials, the electron-phonon interaction matrix elements of $H^{(1)}_{\mathbf{q}j}$
diverge like $\delta_{nn'}/q$, while (ii) the denominator behaves like $q^2/2m^*_{\hat{\mathbf{q}}}$.
In these expressions, $q$ is the norm of the $\mathbf{q}$ wavevector, and $m^*_{\hat{\mathbf{q}}}$ is the effective mass along direction $\mathbf{q}$,
represented by the unit vector $\hat{\mathbf{q}}$, with $\mathbf{q}=q \cdot \hat{\mathbf{q}}$. The effective mass is the
inverse of the second derivative of the eigenenergy with respect to the wavevector along $\hat{\mathbf{q}}$ at the band extremum.

Thus, the integrand at small $q$ behaves like $1/q^4$. Focusing on the ZPR$^{\rm adiab}_{\rm c}$ of the bottom of the conduction band (the handling of
the top of the valence band is similar) one finds, in spherical coordinates, and introducing
a spherical cut-off $q_{\rm c}$ whose limit $q_{\rm c} \rightarrow 0$ has to be taken:
\begin{equation}
{\rm ZPR}^{\rm adiab}_{\rm c}(q_{\rm c}<q)=\int_{4\pi}\int_{q_{\rm c}} \frac{f^{\rm adiab}(q\hat{\mathbf{q}})}{q^2 \big( \frac{q^2}
                                {2m^*_{\hat{\mathbf{q}}}}
+\mathcal{O}(q^3)
\big)} q^2 dq d\hat{\mathbf{q}},
\label{eq:ZPR_c_adiab}
\end{equation}
where $f^{\rm adiab}(q\hat{\mathbf{q}})$ is a smooth function of $q$ (it does not diverge, but also does not tend to zero for $q\rightarrow0$).
As announced~\cite{Ponce2015}, ${\rm ZPR}^{\rm adiab}_{\rm c}(q_{\rm c}<q)$ diverges like $1/q_{\rm c}$ when $q_{\rm c} \rightarrow 0$.

By contrast, if phonon frequencies are not suppressed in Eq.~(8),
which is the non-adiabatic case, one falls back to an integral of the type
(here considered for one isotropic phonon branch only, and one
non-degenerate isotropic electronic band):
\begin{eqnarray}
{\rm ZPR}^{\rm non-adiab}_{\rm c}&&(q_{\rm c}<q)
=
\nonumber
\\
\int_{4\pi}\int_{q_{\rm c}}&&
\frac{f^{\rm non-adiab}(q\hat{\mathbf{q}})}
        {q^2 \big( \frac{q^2}
                                {2m^*_{\hat{\mathbf{q}}}}
                        -\omega_{0}
                       +\mathcal{O}(q^2)
                \big)
         }
 q^2 dq  d\hat{\mathbf{q}},
\label{eq:ZPR_c_nonadiab}
\end{eqnarray}
where the $\mathcal{O}(q^2)$ contribution in the denominator comes from the curvature of $\omega_{\textbf{q}}$ with respect to $q$,
usually much smaller than the curvature of the electronic eigenenergies governed by $m^*_{\hat{\mathbf{q}}}$, explicitly mentioned in Eq.~(\ref{eq:ZPR_c_nonadiab}).
${\rm ZPR}^{\rm non-adiab}_{\rm c}(q_{\rm c}<q)$ does not diverge for $q_{\rm c} \rightarrow 0$, provided $\omega_{0}$ is non-zero.

Acoustic modes should not be forgotten. Their eigenfrequency  $\omega_{0}$ tends to zero in the $q \rightarrow 0$ limit.
However, the corresponding first-order
electron-phonon matrix element does not
diverge like $1/q$, and also,
it has been shown by Allen and Heine~\cite{Allen1976} that, thanks to translational symmetry, the contribution to the Fan term from acoustic modes
is exactly cancelled by the Debye-Waller term, in the $q \rightarrow 0$ limit. Thus no divergence arises from them.

To summarize, finite phonon frequencies at $q=0$ remove the AHC divergence in the non-adiabatic case.


\subsection{The generalized Fr\"ohlich Hamiltonian}
\label{sec:gFr}

Instead of suppressing phonon frequencies appearing in the Fan self-energy Eq.(8), like in the adiabatic approximation,
a radically different strategy is followed
in the generalized Fr\"ohlich Hamiltonian:
one uses only parameters at $q=0$, including crucially the phonon frequencies, and extends them to the whole
$\mathbf{q}$-space, using simple behavior for medium and large-$q$
(i.e. parabolic electronic dispersion and no phonon dispersion).
This corresponds to macroscopic electrostatic treatment and a continuum space hypotheses.

The model proposed by Fr\"ohlich~\cite{Froehlich1954} in 1954 is the most simplified version of such hypotheses: one non-dispersive LO phonon branch, one electronic parabolic band governed by a single, isotropic effective mass, and electron-phonon interaction originating from the isotropic macroscopic dielectric interaction between the IR-active LO mode
and the charged electronic carrier. No Debye-Waller contribution is included. Spin might be ignored, as only one electron is considered, without spin-phonon interaction.
Explicitly, the Fr\"ohlich Hamiltonian reads~\cite{Frohlich1950, Frohlich1952, Froehlich1954, Feynman1955},
\begin{equation}
\hat{H}^{\rm Fr}=\hat{H}^{\rm Fr}_{\rm el}+\hat{H}^{\rm Fr}_{\rm ph}+\hat{H}^{\rm Fr}_{\rm EPI} \, ,
\label{eq:HFr}
\end{equation}
with
\begin{equation}
\hat{H}^{\rm Fr}_{\rm el}= \sum_{\mathbf{k}} \frac{\mathbf{k}^2}{2m^*}
\hat{c}^+_{\mathbf{k}}\hat{c}_{\mathbf{k}} \, ,
\label{eq:HFrel}
\end{equation}
\begin{equation}
\hat{H}^{\rm Fr}_{\rm ph}= \sum_{\mathbf{q}} \omega_{\rm LO}
\hat{a}^+_{\mathbf{q}}\hat{a}_{\mathbf{q}} \, ,
\label{eq:HFrph}
\end{equation}
\begin{equation}
\hat{H}^{\rm Fr}_{\rm EPI}= \sum_{\mathbf{q},\mathbf{k}}
g^{\rm Fr}(\mathbf{q})
\hat{c}^+_{\mathbf{k}+\mathbf{q}}\hat{c}_{\mathbf{k}}
(\hat{a}_{\mathbf{q}}+\hat{a}^+_{-\mathbf{q}}) \, ,
\label{eq:HFrEPI}
\end{equation}
where $\hat{a}^+_{\mathbf{q}}$ and $\hat{a}_{\mathbf{q}}$ are phonon creation and annihilation operators,
$\hat{c}^+_{\mathbf{q}}$ and $\hat{c}_{\mathbf{q}}$
are electron creation and annihilation operators,
and the electron-phonon coupling parameter is given by
\begin{equation}
g^{\rm Fr}(\mathbf{q})=
\frac{i}{q}
\Biggl[ \frac{2 \pi \omega_{\rm LO}}{V_{\rm BvK}}
\Big( \frac{1}{\epsilon^\infty}-\frac{1}{\epsilon^0} \Big)\Biggr]^{1/2},
\label{eq:gFr}
\end{equation}
where $V_{\rm BvK}$ is the Born-von Karman normalisation volume corresponding to the $\mathbf{k}$ and $\mathbf{q}$ samplings,
$q$ is the norm of $\mathbf{q}$ and
$\hat{\mathbf{q}}$ is its direction, $\mathbf{q}=q \cdot \hat{\mathbf{q}}$,
$\epsilon^\infty$ is the optical dielectric constant,
and
$\epsilon^0$ is the low-frequency dielectric constant.
The sums over $\mathbf{k}$ and $\mathbf{q}$ run over the
whole reciprocal space.

At the lowest order of perturbation theory, this yields the well-known result~\cite{Froehlich1954,Feynman1955,Mahan2000} for the polaron binding energy
(that we denote ZPR$^{\rm Fr}$):
\begin{equation}
{\rm ZPR}^{\rm Fr}=-\alpha \omega_{\rm LO}, \,{\rm with}  \, \alpha=
\Big( \frac{1}{\epsilon^\infty}-\frac{1}{\epsilon^0} \Big)
\sqrt{\frac{m^\ast}{2\omega_{\rm LO}}}.
\label{eq:alpha}
\end{equation}
Accurate diagrammatic Monte Carlo simulations~\cite{Mishchenko2000} have demonstrated reasonable validity of this lowest-order perturbation
result in a range
that extends to about  $\alpha \approx 8$. Beyond this, electronic self-trapping becomes too important, and the lowest-order result deviates strongly from the exact results.

Multiphonon generalisations of the Fr\"ohlich model Hamiltonian have been considered
in several occasions, but only in the cubic case, while the anisotropy
of the electronic effective mass has been accounted for in an approximate way,
and not altogether with the degeneracy of band edges, that has either been treated approximately, or for a spherical model, without warping, or only considering one phonon branch see e.g. Refs.~\onlinecite{Mahan1965, Trebin1975, Devreese2010,Nery2016, Schlipf2018,Bhandari2018}.

Our generalized Fr\"ohlich model includes {\it all the features of real materials altogether: multiphonon, anisotropic, degenerate band extrema}. It is derived by considering the full
Eq.~(8)
and applying the same ansatz as the original Fr\"ohlich model:
we treat as exactly as possible the region in the Brillouin
zone where the phonon frequency is large with respect to the eigenenergy differences, and,
coherently, suppress the contributions from bands with a large energy difference with respect to phonon
frequencies (interband contributions), keeping only intraband contributions.
Like for the Fr\"ohlich Hamiltonian, we use only parameters that describe the functions in the Brillouin zone around $q=0$, and extend them to the whole reciprocal space, using the same simple behaviors for medium and large-$q$: parabolic electronic dispersion
and no phonon dispersion.
We also retain only the
macroscopic Fr\"ohlich electron-phonon interaction, $\mathbf{G}=0$ component, following Vogl~\cite{Vogl1976}.
This is also coherent
with the first-principles Fr\"ohlich vertex of Ref.~\onlinecite{Verdi2015},
and the associated macroscopic electric field, derived in the appendix A of Ponc\'e {\it et al.}, Ref.~\onlinecite{Ponce2015}.
The possible degeneracy of the electronic states, causing the warping of the electronic structure, is addressed
following Refs.~\onlinecite{Mecholsky2014} and ~\onlinecite{Laflamme2016}.
It has been shown also by AHC that the contribution to the Fan term from acoustic modes, for which the frequency vanishes when $q \rightarrow 0$,
is exactly cancelled by the Debye-Waller term, close to $q=0$.

Having highlighted the basic ideas of the
derivation of the generalized Fr\"ohlich Hamiltonian, in what follows, we write down the mathematical expressions. We focus on the conduction band expression, ZPR$_{\rm c}$, associated to one of the possibly degenerate
states, labelled $c$, with energy $\varepsilon_{\rm c}$ and with wavevector $\mathbf{k}_0$. Similar expressions for ZPR$_{\rm v}$ can be derived.
For the sake of simplicity, we will
suppose that $\mathbf{k}_0$ is the Brillouin zone center
$\Gamma$, but the equations that follow
can be straightforwardly adapted to another value of
$\mathbf{k}_0$. We also introduce
$\mathbf{k}'=\mathbf{k}+\mathbf{q}$.

Under the above hypotheses, the matrix elements of $H^{(1)}_{\mathbf{q}j}$ for $q \rightarrow 0$
writes
\begin{eqnarray}
\langle \mathbf{k}',n'|H^{(1)}_{\mathbf{q}j}|\mathbf{k},n\rangle
=
\frac{i}{q} &&\frac{4\pi}{\Omega_0}
\Big(  \frac{1}{2\omega_{j0}(\hat{\mathbf{q}})V_{\rm BvK}   }
\Big)^{1/2}
\frac{\hat{\mathbf{q}}.\mathbf{p}_j(\hat{\mathbf{q}})}
        {\epsilon^\infty(\hat{\mathbf{q}})}
\nonumber
\\
\times &&\langle \mathbf{k}',n'|\mathbf{k},n\rangle_{\rm P}
\, ,
\label{eq:SM_H1}
\end{eqnarray}
with $\Omega_0$ the primitive cell volume,
$V_{\rm BvK}$ the Born-von-Karman normalisation volume corresponding to the $\mathbf{k}$ and $\mathbf{q}$ samplings,
and the scalar product
$\langle \mathbf{k}',n'|\mathbf{k},n\rangle_{\rm P}$
is computed from the periodic part of the Bloch functions. Sums over $\mathbf{k}$ and $\mathbf{q}$
of the type
$\sum_{\mathbf{q}}f(\mathbf{q})/V_{\rm BvK}$    are to be replaced by
$\Omega_0/(2 \pi)^3 \int d^3 \mathbf{q} f(\mathbf{q})  $
in the macroscopic limit in what follows.
We have used the notation $\omega_{j0}(\hat{\mathbf{q}})$ for the $q \rightarrow 0$
limit of the $j$-phonon branch
($\omega_{j0}(\hat{\mathbf{q}})=
\lim_{q \rightarrow 0}\omega_{j(q\hat{\mathbf{q}})}$),
while
the related mode-polarity vectors $\mathbf{p}_j(\hat{\mathbf{q}})$, see Eq.~(41) of Ref.~\onlinecite{Veithen2005},
are defined from the Born effective charges and the eigendisplacements of the phonon mode,
summing over the atoms labelled by $\kappa$, and the Cartesian direction of displacement $\alpha$,
\begin{eqnarray}
p_{\gamma,j}(\hat{\mathbf{q}})=
\lim_{q \rightarrow 0}
\sum_{\kappa\alpha}
Z^*_{\kappa\alpha,\gamma}
U_{\kappa\alpha,j}(q\hat{\mathbf{q}}).
\label{eq:SM_modepolarity}
\end{eqnarray}
Similarly, the optic dielectric constant along $\hat{\mathbf{q}}$ is obtained from
the optic dielectric tensor by summing over wavevector components,
$\epsilon^\infty(\hat{\mathbf{q}})=
\sum_{\alpha\beta}\hat{q}_{\alpha}\epsilon^\infty_{\alpha\beta}\hat{q}_{\beta}$,
as in Eq.~(56) of Ref.~\onlinecite{Gonze1997}.
The mode-polarity vector vanishes for all non-IR-active modes. Thus naturally only the LO modes contribute
to the ZPR$^{\rm gFr}$ in this generalized Fr\" ohlich approach (for acoustic modes and TO modes, the mode-polarity
vector vanishes). Still,
the directional dependency of the phonon frequencies, of the mode-polarity vectors,
and of the dielectric constant, is correctly taken into account.

We then restrict ourselves to the bands that are
degenerate with $c$, and choose a $\hat{\mathbf{k}}$-independent complete basis for these states
at $\mathbf{k}_0=\Gamma$, denoted $|\Gamma,m\rangle$.
We also define the
$\hat{\mathbf{k}}$-dependent overlap between states $m$ at $\Gamma$ and states $n$
along $\hat{\mathbf{k}}$ in the $k \rightarrow 0$ limit:
\begin{eqnarray}
s_{nm}(\hat{\mathbf{k}})=
\lim_{k \rightarrow 0}
\langle
k\hat{\mathbf{k}},n|\Gamma,m\rangle_{\rm P} \, .
\end{eqnarray}
This definition supposes that the phase of the
$|k\hat{\mathbf{k}},n\rangle$
states
is chosen continuously as a function of $k$, for vanishing
$k$ along the $\hat{\mathbf{k}}$ direction. The phase can always be chosen to be continuous, and this choice will have no bearing on the results obtained later.
Still, it does not imply any continuity hypothesis for the phase between states $n$ for different $\hat{\mathbf{k}}$ directions.
The matrix $s_{nm}(\hat{\mathbf{k}})$ is unitary.
Note that the $|\Gamma,m\rangle$ states are fixed before the
$\hat{\mathbf{k}}$ direction is known.
The coefficients
$s_{nm}(\hat{\mathbf{k}})$ can be computed from
first principles following the
perturbative treatment of Refs.~\onlinecite{Mecholsky2014} and ~\onlinecite{Laflamme2016}, which is the first-principles
equivalent of the $\mathbf{k}.\mathbf{p}$ approach, restricted
to the subspace spanned by the bands that connect to the degenerate states.
It delivers Luttinger-Kohn parameters
$D_{jj'}^{\alpha\beta}$,~\cite{Luttinger1955} that allows one to find $s_{nm}(\hat{\mathbf{k}})$ for all $\hat{\mathbf{k}}$. The Luttinger-Kohn
parameters can also be determined from experimental measurements.
The $s$ matrices can be
thought as being the adaptation of spherical harmonics to the specific symmetry representation to which the degenerate wavefunctions belong.

With these definitions, we write the generalized Fr\"ohlich electron-phonon interaction as
\begin{eqnarray}
g^{\rm gFr}(\mathbf{q}j,\mathbf{k}n'n)=
\frac{i}{q} &&\frac{4\pi}{\Omega_0}
\Big(  \frac{1}{2\omega_{j0}(\hat{\mathbf{q}})V_{\rm BvK}   }
\Big)^{1/2}
\frac{\hat{\mathbf{q}}.\mathbf{p}_j(\hat{\mathbf{q}})}
        {\epsilon^\infty(\hat{\mathbf{q}})}
\nonumber
\\
\times &&
\sum_m
s_{n'm}(\hat{\mathbf{k}}')
(s_{nm}(\hat{\mathbf{k}}))^*.
\label{eq:SM_ggFr}
\end{eqnarray}
This expression depends on the directions
$\hat{\mathbf{k}}$,
$\hat{\mathbf{k}}'$,
and
$\hat{\mathbf{q}}$, but not explicitly on their norm, except for the $\frac{1}{q}$ factor.

We focus then on the denominators appearing in the self-energy expression Eq.~(8),
 evaluated at
$\omega=\varepsilon_{\rm c}$, following Eq.~(9).
Again we consider only the intraband
terms, for bands $n'$ degenerate with $c$ in the $q \rightarrow 0$ limit. These bands, similarly to the $c$ state,
are unoccupied, hence
$f_{\mathbf{k}'n'}=0$. Thus the numerator of the second term in Eq.~(8) vanishes. We find:
\begin{eqnarray}
\lim_{q \rightarrow 0} (\varepsilon_{\rm c}-\varepsilon_{\mathbf{k}'n'}-
\omega_{\mathbf{q}j})
&=&
-\frac{q^2}{2m^*_{n'}(\hat{\mathbf{k}}')}
-\omega_{j0}(\hat{\mathbf{q}}).
\label{eq:SM_denFan}
\end{eqnarray}
See Refs.~\onlinecite{Mecholsky2014} and ~\onlinecite{Laflamme2016}
for the detailed treatment of $m^*_{n'}(\hat{\mathbf{k}}')$, the directional effective mass associated to one electronic band among
a set of degenerate bands at an extremum.

Having clarified the electronic, phononic and electron-phonon interaction ingredients, we can write the corresponding Hamiltonian,
Eqs.(1)-(5) that indeed generalizes properly Eqs.~(\ref{eq:HFr}-\ref{eq:gFr}).


\subsection{Lowest-order perturbation treatment}
\label{sec:perturbative}

Given the Hamiltonian, the lowest-order perturbation
ZPR value can be inferred by integration
over the radial $q$ coordinate.
One obtains for the conduction bands:
\begin{eqnarray}
{\rm ZPR}_{\rm c}^{\rm gFr}&=&
-\frac{1}{\pi\Omega_0} \int_{4\pi} d\hat{\mathbf{q}}
\sum_{jn'}
\frac{|s_{n'c}(\hat{\mathbf{q}})|^2} {2\omega_{j0}(\hat{\mathbf{q}})   }
\nonumber
\\
&\times
&\Big(
\frac{ \hat{\mathbf{q}}.\mathbf{p}_j(\hat{\mathbf{q}}) }
        { \epsilon^\infty(\hat{\mathbf{q}}) }
\Big)^2
\Big(\frac{2m^*_{n'}(\hat{\mathbf{q}})}{\omega_{j0}(\hat{\mathbf{q}})}\Big)^{1/2}\frac{\pi}{2}.
\label{eq:SM_ZPR_2Dintegral}
\end{eqnarray}

Let us denote by $n_{\rm deg}$ the degeneracy of the band edge.
In the non-degenerate case ($n_{\rm deg}=1$), only $n'={\rm c}$ has to be considered, then obviously
$|s_{n'{\rm c}}(\hat{\mathbf{q}})|=1$
independently of $\hat{\mathbf{q}}$.
Grouping terms adequately, we obtain
\begin{eqnarray}
{\rm ZPR}_{\rm c}^{\rm gFr}&=&
-\sum_{j}
\frac{1}{\sqrt{2}\Omega_0}
\int_{4\pi} d\hat{\mathbf{q}}
\big( m_{\rm c}^*(\hat{\mathbf{q}})
\big)^{1/2}
\nonumber
\\
&\times &\big(\omega_{j0}(\hat{\mathbf{q}})
\big)^{-3/2}
\Big(
\frac{\hat{\mathbf{q}}.\mathbf{p}_j(\hat{\mathbf{q}})}
        {\epsilon^\infty(\hat{\mathbf{q}})}
\Big)^2.
\label{eq:ZPR_c_Fr_nondeg}
\end{eqnarray}
The link can then be established with Eq.~(\ref{eq:alpha}) in case of isotropic effective mass, isotropic dielectric tensor, and single LO phonon branch.

Let us further examine the degenerate case ($n_{\rm deg} > 1 $).
We suppose that the degeneracy is not accidental, but due to symmetry reasons.
In this case, we can exploit the symmetry
to show that the computation $|s_{n'{\rm c}}(\hat{\mathbf{q}})|^2$ can be avoided,
as this factor can be replaced by
$1/n_{\rm deg}$
in the integral. We will present the demonstration for the case of
a three-fold degeneracy, that arises due to a cubic space group.
This demonstration could be made general thanks to group theory.

At the bottom of the conduction band, located at the $\mathbf{k}_0$ wavevector, but that for convenience we again choose to be $\Gamma$, we set up a
basis of three orthonormalized eigenfunctions, denoted $\{|X\rangle,|Y\rangle,|Z\rangle\}$, that form
an irreducible representation of the symmetry group.
The eigenfunction $|\Gamma,{\rm c}\rangle$
for which we compute the ZPR$^{\rm gFr}_{\rm g}$, can be decomposed in this basis, with coefficients $(u_{{\rm c}X},u_{{\rm c}Y},u_{{\rm c}Z})$:
\begin{eqnarray}
|\Gamma,{\rm c}\rangle=u_{{\rm c}X}|X\rangle+u_{{\rm c}Y}|Y\rangle+u_{{\rm c}Z}|Z\rangle.
\label{eq:SM_psic}
\end{eqnarray}
The eigenfunctions $|q\hat{\mathbf{q}},n'\rangle$ over which the $n'$ sum and $\hat{\mathbf{q}}$
integral are done in Eq.~(\ref{eq:SM_ZPR_2Dintegral}) can be decomposed as well in this basis:
\begin{eqnarray}
|q\hat{\mathbf{q}},n'\rangle=s_{n'X}(\hat{\mathbf{q}})|X\rangle+s_{n'Y}(\hat{\mathbf{q}})|Y\rangle+s_{n'Z}(\hat{\mathbf{q}})|Z\rangle
\nonumber
\\
\label{eq:SM_psi_n_prime}
\end{eqnarray}
Thus Eq.~(\ref{eq:SM_ZPR_2Dintegral}) becomes
\begin{eqnarray}
{\rm ZPR}_{\rm c}^{\rm gFr}=
\sum_{m \in \{X,Y,Z\}}
\sum_{m' \in \{X,Y,Z\}}
u_{{\rm c}m}^* T_{mm'} u_{{\rm c}m'}
\label{eq:SM_ZPR_c_deg}
\end{eqnarray}
where
\begin{eqnarray}
T_{mm'}=
-\sum_{jn'}
\frac{1}{\sqrt{2}\Omega_0}
&&
\int_{4\pi} d\hat{\mathbf{q}}
s_{n'm}(\hat{\mathbf{q}})  s_{n'm'}^*(\hat{\mathbf{q}})
\big( m_{n'}^*(\hat{\mathbf{q}})
\big)^{1/2}
\nonumber
\\
&&\big(\omega_{j0}(\hat{\mathbf{q}})
\big)^{-3/2}
\Big(
\frac{\hat{\mathbf{q}}.\mathbf{p}_j(\hat{\mathbf{q}})}
        {\epsilon^\infty(\hat{\mathbf{q}})}
\Big)^2.
\label{eq:SM_T}
\end{eqnarray}
The second-rank tensor $T_{mm'}$ does not depend on the state $c$. However the same ZPR$^{\rm gFr}_{\rm c}$ must be obtained
from Eq.~(\ref{eq:SM_ZPR_c_deg})
for any symmetrically equivalent
state $c'$. Hence, the tensor $T_{mm'}$ must be invariant
under all the symmetry operations of the group, which means
\begin{eqnarray}
T_{mm'}=t \delta_{mm'},
\label{eq:SM_T_multipleofunity}
\end{eqnarray}
where $\delta$ is the Kronecker symbol. $t$ can be evaluated thanks to the trace of $T_{mm'}$,
\begin{eqnarray}
n_{\rm deg} \cdot t=
&&
-\sum_{jn'}
\frac{1}{\sqrt{2}\Omega_0}
\int_{4\pi} d\hat{\mathbf{q}}
\Bigg(\sum_{m \in \{X,Y,Z\}} s_{n'm}(\hat{\mathbf{q}})  s_{n'm}^*(\hat{\mathbf{q}})\Bigg)
\nonumber
\\
&&
\big( m_{n'}^*(\hat{\mathbf{q}})
\big)^{1/2}\big(\omega_{j0}(\hat{\mathbf{q}})
\big)^{-3/2}
\Big(
\frac{\hat{\mathbf{q}}.\mathbf{p}_j(\hat{\mathbf{q}})}
        {\epsilon^\infty(\hat{\mathbf{q}})}
\Big)^2,
\label{eq:SM_t}
\end{eqnarray}
yielding
\begin{eqnarray}
{\rm ZPR}_{\rm c}^{\rm gFr}&=&
-\sum_{jn}
\frac{1}{\sqrt{2}\Omega_0
n_{\rm deg}}
\int_{4\pi} d\hat{\mathbf{q}}
\big( m_n^*(\hat{\mathbf{q}})
\big)^{1/2}
\nonumber
\\
&\times &\big(\omega_{j0}(\hat{\mathbf{q}})
\big)^{-3/2}
\Big(
\frac{\hat{\mathbf{q}}.\mathbf{p}_j(\hat{\mathbf{q}})}
        {\epsilon^\infty(\hat{\mathbf{q}})}
\Big)^2,
\label{eq:ZPR_c_Fr}
\end{eqnarray}
since $s_{nm}(\hat{\mathbf{k}})$ is unitary.
Similar equations can be derived for the top of the valence band, yielding
${\rm ZPR}_{\rm v}^{\rm gFr}$.

\subsection{Generalized Fr\"ohlich model with parameters
from first-principles calculations}
\label{sec:gFr_FP}

\begin{table}[h]
\renewcommand\thetable{S5}
\caption{\label{tab:FROvsFP}
For thirty materials: fundamental gap (E$_{\rm g}^{\rm exp}$);
location and possible degeneracy of the band edges (valence/conduction)
from Ref.\onlinecite{MP} except for SrTiO$_3$;
first-principles renormalization of the band edges
(valence band maximum  ZPR$_{\rm v}$, conduction band minimum
ZPR$_{\rm c}$) and of the band gap
(ZPR$_{\rm g}$); and ratio between ZPR$_{\rm g}$ and E$_{\rm g}$ ($R_{\rm g}$).
Methodology: ``exp" for experimental results~\cite{note-experimentalgap}, ``gFr" for generalized Fr\"ohlich model results, Eq.~(37),
and ``AHC" for AHC first-principles results
(with correction Eq.~(3).
}
\begin{tabular}{ | l | r | c | r | r | r | r | r | r |}
 \hline \hline
   & E$_{\rm g}$ & Edge & ZPR$_{\rm v}$ & ZPR$_{\rm v}$ & ZPR$_{\rm c}$ &  ZPR$_{\rm c}$ & ZPR$_{\rm g}$ & $R_{\rm g}$ \\
   & exp    & v/c & gFr           & AHC       & gFr           & AHC        & AHC        &AHC \\
 Material   & eV & deg. & meV & meV & meV & meV & meV & \% \\
 \hline \hline
Ge-dia     &  0.74 &$\Gamma$/L&   0 &  17 &    0 &  -16 &  -33 &  -4.5 \\
Si-dia     &  1.17 &$\Gamma$(3)/&   0 &  35 &    0 &  -21 &  -56 &  -4.8 \\
      &    & X*(X-$\Gamma$)&    &  &    &  &  & \\
GaAs-zb    &  1.52 &$\Gamma$(3)/$\Gamma$&   3 &  19 &   -1 &   1 & -18 &  -1.2 \\
CdTe-zb    &  1.61 &$\Gamma$(3)/$\Gamma$&  12 &  17 &   -4 &   -3 &  -20 &  -1.2 \\
AlSb-zb   &  1.69 &  $\Gamma$(3)/  & 4 & 24 & -4 &-27 &-51 & -3.0   \\
      &    & X*(X-$\Gamma$)&    &  &    &  &  & \\
CdSe-zb    &  1.85 &$\Gamma$(3)/$\Gamma$&  20 &  26 &   -6 &   -8 &  -34 &  -1.8 \\
AlAs-zb    &  2.23 &$\Gamma$(3)/X&  12 &  39 &   -9 &  -35 &  -74 &  -3.3 \\
GaP-zb   &  2.34 & $\Gamma$(3)/   & 9 &28 &-7 & -37 & -65 & 2.8 \\
      &    & X*(X-$\Gamma$)&    &  &    &  &  & \\
ZnTe-zb    &  2.39 &$\Gamma$(3)/$\Gamma$&  11 &  19 &   -4 &   -3 &  -22 &  -0.9 \\
SiC-zb     &  2.40 &$\Gamma$(3)/X&  59 & 112 &  -32 &  -67 & -179 &  -7.5 \\
CdS-zb     &  2.42 &$\Gamma$(3)/$\Gamma$&  39 &  44 &  -15 &  -26 &  -70 &  -2.9 \\
AlP-zb     &  2.45 &$\Gamma$(3)/X&  21 &  54 &  -14 &  -39 &  -93 &  -3.8 \\
ZnSe-zb    &  2.82 &$\Gamma$(3)/$\Gamma$&  22 &  31 &   -8 &  -13 &  -44 &  -1.6 \\
TiO$_2$-t  &  3.03 &$\Gamma$/$\Gamma$& 236 & 195 & -135 & -142 & -337 & -11.1 \\
SrTiO$_3$ &3.25 &R(3) / & 140 & 191 & -115 &  -99 & -290 &  -8.9 \\
      &    & $\Gamma$(3) \cite{Benrekia2012}&    &  &    &  &  & \\
GaN-w      &  3.20 &$\Gamma$/$\Gamma$&  80 & 101 &  -31 &  -88 & -189 &  -5.9 \\
GaN-zb     &  3.29 &$\Gamma$/$\Gamma$&  77 &  85 &  -30 &  -91 & -176 &  -5.3 \\
ZnO-w      &  3.44 &$\Gamma$/$\Gamma$& 106 & 108 &  -38 &  -49 & -157 &  -4.6 \\
SnO$_2$-t  &  3.73 &$\Gamma$/$\Gamma$& 144 & 176 &  -47 &  -39 & -215 &  -5.8 \\
ZnS-zb     &  3.91 &$\Gamma$/$\Gamma$&  40 &  49 &  -19 &  -39 &  -88 &  -2.3 \\
BaO-rs     &  4.80 &X/X& 225 & 189 & -133 &  -82 & -271 &  -5.6 \\
SrO-rs     &  5.22 &$\Gamma$(3)/X& 229 & 214 & -141 & -112 & -326 &  -6.2 \\
C-dia      &  5.48 &$\Gamma$(3) / &   0 & 134 &    0 & -196 & -330 &  -6.0 \\
      &    & X*(X-$\Gamma$)&    &  &    &  &  & \\
AlN-w      &  6.20 &$\Gamma$/$\Gamma$& 151 & 204 &  -80 & -195 & -399 &  -6.4 \\
BN-zb      &  6.25 &$\Gamma$(3)/X&  94 & 201 &  -68 & -205 & -406 &  -6.5 \\
CaO-rs     &  6.93 &$\Gamma$(3)/X& 223 & 211 & -154 & -130 & -341 &  -4.9 \\
Li$_2$O    &  7.55 &$\Gamma$(3)/X& 364 & 347 & -172 & -226 & -573 &  -7.6 \\
MgO-rs     &  7.83 &$\Gamma$(3)/$\Gamma$& 327 & 317 & -137 & -207 & -524 &  -6.7 \\
SiO$_2$-t  &  8.90 &$\Gamma$/$\Gamma$& 292 & 300 & -167 & -285 & -585 &  -6.6 \\
BeO-w      & 10.59 &$\Gamma$(2)/$\Gamma$& 289 & 369 & -191 & -330 & -699 &  -6.6 \\
\hline \hline
\end{tabular}
\end{table}

Table S5
presents the ZPR for the bottom of the conduction band and the top of the
valence band from the present first-principles calculations: either based on the AHC approach,
or from feeding the few relevant first-principles parameters in the generalized Fr\"ohlich model.
These data are used in Fig. 4.

\begin{table}[h]
\renewcommand\thetable{S6}
\caption{\label{tab:SMgFrparameters}
First-principles GGA-PBE parameters for the computation of the ZPR of band edges using the (generalized) Fr\"ohlich model. Only materials with one IR-active phonon branch,
cubic space group, and non-degenerate band extremum (though possibly anisotropic effective masses)
are included in this table.
For AlSb and GaP, the edge is located close to the X point, along the X-$\Gamma$ line, which is denoted X*.
}
\begin{tabular}{ | l | l | r | r | r | r |  r | r | r |}
 \hline \hline
                 & edge & $\epsilon^{\infty}$ & $\epsilon^{0}$ & $\omega_{\rm LO}$ &  $m^*_{xx}$ & $m^*_{zz}$  & $\alpha$ & ZPR$^{\rm gFr}$ \\
 Material   &          &                               &                         & (meV)                  &   $m^*_{yy}$  &               &            & (meV) \\
 \hline \hline
GaAs-zb & c $\Gamma$  & 15.31 & 17.55 & 33.5 & 0.009 & 0.009& 0.016 &-1 \\
CdTe-zb & c $\Gamma$  & 8.89 & 12.37 & 19.1 & 0.052 & 0.052 & 0.192 &-4 \\
AlSb-zb & c X*  & 12.02 & 13.35 & 39.8 & 0.222 & 1.142 &0.090  & -4\\
CdSe-zb & c $\Gamma$  & 7.83 & 11.78 & 23.6 & 0.051 & 0.051 & 0.233 &-6 \\
AlAs-zb & c X                  & 9.49 & 11.51 & 47.3 & 0.243 & 0.897 & 0.184 &-9   \\
GaP-zb & c X*  & 10.50 & 12.53 & 48.6  & 0.230  & 1.062  & 0.152 & -7\\
ZnTe-zb & c $\Gamma$  & 9.05 & 11.99 & 24.1 & 0.076 & 0.076 & 0.178 &-4   \\
SiC-zb & c X                   & 6.97 & 10.30 & 117.0 & 0.228 & 0.677 & 0.280 &-32 \\
CdS-zb & c $\Gamma$  & 6.21 & 10.24 & 34.4 & 0.118 & 0.118 & 0.432 &-15 \\
AlP-zb & c X                   & 8.12 & 10.32 & 59.9 & 0.252 & 0.809 & 0.184 &-14  \\
ZnSe-zb & c $\Gamma$  & 7.35 & 10.73 & 29.3 & 0.089 & 0.089 & 0.276 &-8   \\
GaN-zb & c $\Gamma$  & 6.13 & 11.00 & 86.0 & 0.144 & 0.144 & 0.345 &-30 \\
ZnS-zb & c $\Gamma$  & 5.97 & 9.40 & 40.6 & 0.167 & 0.167 & 0.458 &-19   \\
BaO-rs & v X  & 4.21 & 92.43 & 47.3 & 4.035 & 0.431 & 4.757 & 225   \\
BaO-rs & c X  & 4.21 & 92.43 & 47.3 & 0.380 & 1.197 & 2.812 & -132  \\
SrO-rs & c X   & 3.77 & 20.91 & 55.4 & 0.407 & 1.225  & 2.545 & -141 \\
BN-zb & c X  & 4.52 & 6.69 & 161.0 & 0.299 & 0.895 & 0.422 & -68 \\
CaO-rs & c X  & 3.77 & 16.76 & 66.8 & 0.443 & 1.424 & 2.305 & -154  \\
Li$_2$O & c X  & 2.90 & 7.80 & 86.3 & 0.437 & 0.850 & 1.993 & -172 \\
MgO-rs & c $\Gamma$  & 3.23 & 11.14 & 84.5 & 0.340 & 0.340 & 1.624 & -137 \\
\hline \hline
\end{tabular}
\end{table}

Table S6
presents the parameters computed from first principles,
fed in the (generalized) Fr\"ohlich model, only for materials with one phonon branch,
with cubic symmetry and edges with non-degenerate electronic state, though
including both isotropic or anisotropic effective masses.
The parameters for
other materials will be reported elsewhere.
Most of such edges are found at the bottom of the conduction band, as most
top of valence bands are degenerate.
The information in
Table VI
allows one to compute
the Fr\"ohlich $\alpha$ parameter for these edges,
ranging from 0.016 to 4.757.
The lowest-order perturbation results
are rather accurate in this range. A similar parameter to measure the breakdown
of the perturbation theory in the case of degenerate electronic states is still to be established
to our knowledge.

\subsection{Generalized Fr\"ohlich model with parameters
from experimental data}
\label{sec:gFr_Exp}

We discuss how a few-parameter Fr\"ohlich Hamiltonian, of the type we have introduced, can be obtained from experimental data, thus constituting a few-parameter model of the ZPR of real materials.

In the Fr\"ohlich model~\cite{Frohlich1950,Frohlich1952,Froehlich1954,Feynman1955} four parameters are needed: the LO phonon frequency $\omega_{\rm LO}$, the electronic dielectric constant $\epsilon^\infty$, the low-frequency dielectric constant $\epsilon^0$, and the effective mass $m^\ast$.
These four parameters combine to give the $\alpha$ parameter mentioned in Eq.~(\ref{eq:gFr}). The Fr\"ohlich Hamiltonian can be expressed in the natural units of electronic effective mass and phonon frequency, to make appear $\alpha$ as the single parameter defining the Fr\"ohlich Hamiltonian ~\cite{Frohlich1950, Mahan2000}.
Despite its simplicity, this Fr\"ohlich Hamiltonian has resisted analytical treatments.

One can consider the generalization of such Fr\"ohlich Hamiltonian to arbitrary electronic dispersion, phonon dispersion or electron-phonon interaction, as mentioned in
the introduction of Ref.~\onlinecite{Devreese2009}.
However, one has a model for a real material only if the parameters
describing such dispersions and such interaction are available. Ideally, there should be as few parameters as possible if the goal is to extract the relevant physics.

In this respect, the Hamiltonian in Eqs.~(1)-(5) is the proper generalization of Eqs.~(19)-(23) to all band extrema (degenerate or not, isotropic or anisotropic) in all crystal types (cubic or not), because it can be parameterized with only a few numbers, and all such parameters can be found, in principle, from experiment, even if, in the present work, they were deduced from first principles.

Indeed, the electronic part, Eq.~(2), is fully determined if the so-called Luttinger-Kohn parameters~\cite{Luttinger1955} are available.
Taking the example of the top of the valence band of a cubic material, with three-fold degeneracy, only three Luttinger-Kohn parameters, called $A$, $B$, and $C$ in Ref.~\onlinecite{Luttinger1955} are needed to obtain the direction-dependent effective masses,
$m^*_n(\hat{\mathbf{k}})$, of the three electronic bands,
for all directions $\mathbf{k}$. Similarly, the direction-dependent phonon frequencies $\omega_{j0}(\hat{\mathbf{q}})$ and their mode-polarity vectors
$p_{\gamma,j}(\hat{\mathbf{q}})$ can be obtained from neutron and infra-red spectra measurements and parametrized with a few numbers. For a cubic material
with two atoms per unit cell, there is only one LO frequency,
with mode-polarity vector directly determined from the LO-TO splitting.
The optic dielectric constant along any direction $\mathbf{q}$
comes from the knowledge of a 3x3 symmetric tensor, easily obtained by optical measurements.

Conversely, in Ref.~\onlinecite{Verdi2015}, by Verdi and Giustino, a model {\it for the electron-phonon interaction (or vertex)}
is proposed (at variance with our model of the full Hamiltonian, with a simplified vertex).
It is used for a very different purpose.
In this very nice study, Verdi
and Giustino used the Fr\"ohlich idea
and the microscopic theory developed by Vogl~\cite{Vogl1976}
to model the non-analytic behaviour of the electron-phonon vertex
in the $\bf{q} \rightarrow 0$ limit.
This approach was instrumental to make the electron-phonon interaction amenable
to Fourier interpolation
in an attempt to diminish the computational resources needed
to capture the long-range behaviour of the scattering potentials and
perform mobility calculations using Wannier functions.
However, their work does not propose a ZPR model or Fr\"ohlich Hamiltonian per se.
Indeed,  the electronic dispersion in
Verdi and Giustino is the full first-principles dispersion in the entire Brillouin Zone and similarly
for the phonon dispersion, unlike the one of the original
Fr\" ohlich approach (one parameter
for each of these dispersions), and the present one, which relies on few parameters, the number of which depends on the crystal type and symmetries, and on the Luttinger-Kohn parameters.

\section{Zero-point renormalization in the ASC approach.}
\label{sec:ZPR_ASC}

\subsection{ASC Formalism and relation with the AHC}
\label{sec:ASC}

In the adiabatic supercell approach,
the temperature-dependent average band edges (here written for the bottom of the conduction band) are obtained from
\begin{eqnarray}
\langle \varepsilon_{\rm c}(T) \rangle =Z_I ^{-1} \sum_m \exp(- \beta E_m) \langle\varepsilon_{\rm c} \rangle_m,
\label{eq:eigenASC}
\end{eqnarray}
where $\beta=k_BT$ (with $k_B$ the Boltzmann constant), $T$ the temperature, $Z_I$ the canonical partition function among the quantum nuclear states $m$ with energies $E_m$
($Z_I=\sum_m \exp(- \beta E_m)$),
and $\langle\varepsilon_{\rm c} \rangle_m$
the band edge average taken over the corresponding many-body nuclear wavefunction.
See e.g. Eq.~(4) of Ref.~\onlinecite{Monserrat2016b}.

Such approach allows one to perform nonperturbative calculations,
and thus includes effects beyond the adiabatic AHC approach, as emphasized in Sec.XI.A.2
of Ref.~\onlinecite{Giustino2017}. It also allows one to use as starting point the
GW eigenenergies, instead of the DFT ones, as proposed and used by some of us~\cite{Antonius2014},
and further exploited by Monserrat~\cite{Monserrat2016a} and Karsai et al.~\cite{Karsai2018}.

As detailed in Ref.~\onlinecite{Antonius2015}, and illustrated in Fig. 4 of this reference,
one can dissociate the harmonic approximation
for the nuclear positions from the (non)quadratic behaviour of the eigenenergies
as a function of the nuclear positions. While phonon anharmonicities can be safely ignored
for most of the materials in the present context (except SrTiO$_3$,
see the METHODS section of the main text,
the (non)quadratic behaviour might very important in the ASC formalism.
Within the harmonic approximation, one finds Eq.~(19) of Ref.~\onlinecite{Antonius2015}:
\begin{eqnarray}
\langle \varepsilon_{\rm c}(T) \rangle=Z^{-1}_I \sum_{\bf n} \exp(- \beta E[{\bf n}])
\int d{\bf z} |\chi_{\bf n}({\bf z})|^2  \varepsilon_{\rm c}({\bf z}).
\nonumber
\\
\label{eq:eigenASCharmonic}
\end{eqnarray}
In this equation, $\bf z$ denotes the ensemble of all phonon coordinate
amplitudes, while ${\bf n}$ denotes
the ensemble of all phonon occupation numbers.
As such, this expression does not diverge for IR-active materials, thanks to the
sampling over coordinate amplitudes and the $\varepsilon_{\rm c}({\bf z})$ asymptotic linear behaviour, see Fig. 4 of Ref.~\onlinecite{Antonius2015} or Fig. 2 of Ref.~\onlinecite{Monserrat2015}.
By contrast, making the quadratic approximation for the $\varepsilon_{\rm c}({\bf z})$ yields
the AHC result for the Fan term, and hence, a divergence: the curvature becomes indeed
infinite for IR-active longwavelength phonons. An extreme nonquadratic behavior is also observed
in the degenerate case, as illustrated in Fig. 1 of Ref.~\onlinecite{Karsai2018}.
The connection with the adiabatic AHC result has been analyzed in detail in Ref.~\onlinecite{Ponce2014a}.

However, obviously, Eq.~(\ref{eq:eigenASC}) as well as Eq.~(\ref{eq:eigenASCharmonic}) are based on the adiabatic
approximation: the quantum nuclear state is
not responding to the presence of an added or removed electron. Instead, Eq.~(\ref{eq:eigenASC}) should have reflected such dependence to be correct.
The correlation between electron and nuclei is not present in Eq.~(\ref{eq:eigenASC}): the electron
is affected by the phonons, but the phonons are incorrectly not affected by the addition of an electron.
For IR-active materials, as explained in the main text, the lack of such correlation has important
consequences for the predictive capability of the method.

As another consequence of the time-dependent atomic motion, the nonquadratic behaviour, present in the adiabatic treatment, looses its importance.
The behaviour observed for degenerate electronic state,
in the adiabatic approximation, exemplified by Fig. 1 of Ref.~\onlinecite{Karsai2018} might nevertheless still be relevant in the context of the Jahn-Teller effect.
Anyhow, the transition between states, due to the first-order electron-phonon interaction, always involves a modification of the energy by the emission
or absorption of a phonon. Hence, the corresponding initial and final energies are not identical.

\subsection{The non-quadratic effect for IR-inactive materials}
\label{sec:NQ}

Coming back to the C-diamond case, the ``anharmonic effect" in the adiabatic approximation is argued to be large in the above-mentioned Sec.III of Ref.~\onlinecite{Antonius2015}, that some of us co-authored. At present, we call it more properly “non-quadratic effect” following Ref.~\onlinecite{Monserrat2015}.
However, this study was performed within the adiabatic approximation. In the case of infrared active materials, without such “non-quadratic effects”, the ZPR is infinite in the adiabatic approximation.
The non-quadratic effects succeed to bring the adiabatic ZPR to a finite value, i.e. it has “infinite” impact.
For C, without such “non-quadratic effects”, the ZPR is finite in the adiabatic approximation, but nevertheless comes from the integration of a diverging integrand.
Then the non-quadratic effects modify the adiabatic harmonic value by 40\% by suppressing this diverging integrand.
Still, despite yielding a finite value, the adiabatic approximation is incorrect to start with: if atomic motion effects are taken into account (and they should be taken into account from the very start, since the physics is the one of electrons not being able to follow the atomic motions), there is no diverging integrand that the non-quadratic effects would reduce afterwards.
Thus, the crucial role of atom dynamics cannot be ignored even in diamond.
We actually think that the estimation of non-quadratic effects in our publication Ref.~\onlinecite{Antonius2015} or in Ref.~\onlinecite{Monserrat2015} cannot be transposed to the non-adiabatic treatment, even for the case of C.
The non-quadratic effects in Ref.~\onlinecite{Antonius2015} or Ref.~\onlinecite{Monserrat2015} remove a non-existing divergence, either global (IR-active materials), or integrable (non-IR-active materials), spuriously introduced by the adiabatic approximation.

\subsection{GW approximation within the ASC: discussion}
\label{sec:GW}

In such calculation of the ZPR of diamond, Ref.~\onlinecite{Antonius2014}, we have obtained a 40\% change of direct gap ZPR when the quantities needed in the ASC approach were obtained from GW instead of DFT.
This effect appears to be
comparable to the difference between AHC and experimental values reported in the present
manuscript, while (as for the non-quadratic effect) C is an IR-inactive material.

In Ref.~\onlinecite{Karsai2018}, a similar change was reproduced for the direct gap, but the change for the indirect (true) gap of diamond was much smaller, about 23\%.
For the other materials studied in Ref.~\onlinecite{Karsai2018}, using the ASC approach, the ratio between GW and DFT, systematically obtained, was usually closer to unity, but not always.
One can wonder whether such difference between GW and DFT would not also be obtained within the non-adiabatic AHC approach.

The estimation of such an effect within the non-adiabatic treatment is at present out of reach.
Actually, as already argued in the previous sections, one can hardly justify transposing the conclusions obtained in the ASC approach to the non-adiabatic AHC approach, because the ASC approach relies on the non-quadratic coupling to eliminate the divergence of the ZPR, while for most materials, we believe such elimination to originate from the coupled dynamics of electron and phonons.
In view of the agreement obtained in Fig. 2 for non-adiabatic AHC+DFT, we can simply infer that the DFT to GW effect is apparently smaller in the non-adiabatic AHC case than in the ASC case.

Also, let us emphasize that such a DFT to GW effect can hardly equal the non-adiabatic effect reported in the present manuscript.
Indeed, for IR-active materials, such effect suppresses an infinity and replaces it by a finite value, which is far more important than a 40\% modification.
Even if the non-quadratic effect is taken into account, Fig. 2 shows that the difference between ASC and AHC is not merely 40\% for many materials, but can be as large as a factor of 2 or 3.

So, we do not deny that other effects, like GW, could be important at the level of the 25\% agreement that we show in Fig. 2. However, the non-adiabatic effect is not simply a 25\% modification, it reduces a divergent quantity (possibly integrable) to a finite value.

\section*{References}
\bibliography{main}
\end{document}